\documentclass[twocolumn]{aastex63}
\usepackage{amssymb}
\usepackage{textcomp}
\usepackage{subfigure}
\usepackage{appendix}

\usepackage{comment}
\usepackage{rotating}

\newcommand{\x}{\mbox{$\times$}}
\newcommand{\Msun}{\mbox{$M_{\odot}$}}

\received{}
\revised{}
\accepted{}
\submitjournal{ApJ}
\shorttitle{}
\shortauthors{}
\graphicspath{{./}{figures/}}
\begin{document}
\title{The ALMA Survey of 70 $\mu$m Dark High-mass Clumps in Early Stages (ASHES).\\ IV. Star formation signatures in G023.477 
}
\author{Kaho Morii}
\affil{Department of Astronomy, Graduate School of Science, The University of Tokyo, 7-3-1 Hongo, Bunkyo-ku, Tokyo 113-0033, Japan email: kaho.morii@grad.nao.ac.jp}
\affil{National Astronomical Observatory of Japan, National Institutes of Natural Sciences, 2-21-1 Osawa, Mitaka, Tokyo 181-8588, Japan}

\author{Patricio Sanhueza}
\affil{National Astronomical Observatory of Japan, National Institutes of Natural Sciences, 2-21-1 Osawa, Mitaka, Tokyo 181-8588, Japan}
\affil{Department of Astronomical Science, SOKENDAI (The Graduate University for Advanced Studies), 2-21-1 Osawa, Mitaka, Tokyo 181-8588, Japan}

\author{Fumitaka Nakamura}
\affil{National Astronomical Observatory of Japan, National Institutes of Natural Sciences, 2-21-1 Osawa, Mitaka, Tokyo 181-8588, Japan}
\affil{Department of Astronomical Science, SOKENDAI (The Graduate University for Advanced Studies), 2-21-1 Osawa, Mitaka, Tokyo 181-8588, Japan}
\affil{Department of Astronomy, Graduate School of Science, The University of Tokyo, 7-3-1 Hongo, Bunkyo-ku, Tokyo 113-0033, Japan}

\author{James M. Jackson}
\affil{SOFIA Science Center, USRA, NASA Ames Research Center, Moffett Field CA 94045, USA}

\author{Shanghuo Li}
\affiliation{Korea Astronomy and Space Science Institute, 776 Daedeokdae-ro, Yuseong-gu, Daejeon 34055, Republic of Korea}

\author{Henrik Beuther}
\affil{Max Planck Institute for Astronomy, Konigstuhl 17, 69117 Heidelberg, Germany}

\author{Qizhou Zhang}
\affiliation{Center for Astrophysics $|$ Harvard \& Smithsonian, 60 Garden Street, Cambridge, MA 02138, USA}

\author{Siyi Feng}
\affil{Department of Astronomy, Xiamen University, Xiamen, Fujian 361005, P. R. China}

\author{Daniel Tafoya}
\affil{Department of Space, Earth and Environment, Chalmers University of Technology, Onsala Space Observatory, 439~92 Onsala, Sweden}

\author{Andr\'es E. Guzm\'an}\affil{National Astronomical Observatory of Japan, National Institutes of Natural Sciences, 2-21-1 Osawa, Mitaka, Tokyo 181-8588, Japan}

\author{Natsuko Izumi}
\affil{National Astronomical Observatory of Japan, National Institutes of Natural Sciences, 2-21-1 Osawa, Mitaka, Tokyo 181-8588, Japan}
\affil{College of Science, Ibaraki University, 2-1-1 Bunkyo, Mito, Ibaraki 310-8512, Japan}

\author{Takeshi Sakai}
\affil{Graduate School of Informatics and Engineering, The University of Electro-Communications, Chofu, Tokyo 182-8585, Japan}

\author{Xing Lu}\affil{National Astronomical Observatory of Japan, National Institutes of Natural Sciences, 2-21-1 Osawa, Mitaka, Tokyo 181-8588, Japan}

\author{Ken'ichi Tatematsu}\affil{National Astronomical Observatory of Japan, National Institutes of Natural Sciences, 2-21-1 Osawa, Mitaka, Tokyo 181-8588, Japan}

\author{Satoshi Ohashi}
\affil{RIKEN Cluster for Pioneering Research, 2-1, Hirosawa, Wako-shi, Saitama 351-0198, Japan}

\author{Andrea Silva}\affil{National Astronomical Observatory of Japan, National Institutes of Natural Sciences, 2-21-1 Osawa, Mitaka, Tokyo 181-8588, Japan}

\author{Fernando A. Olguin}
\affil{Institute of Astronomy and Department of physics, National Tsing Hua University, Hsinchu 30013, Taiwan}

\author{Yanett Contreras}
\affil{Leiden Observatory, Leiden University, PO Box 9513, NL-2300 RA Leiden, the Netherlands}

\begin{abstract}
With a mass of $\sim$1000\,$M_\odot$ and a surface density of $\sim$0.5\,g\,cm$^{-2}$, G023.477+0.114 also known as IRDC 18310-4 is an infrared dark cloud (IRDC) that has the potential to form high-mass stars and has been recognized as a promising prestellar clump candidate. 
To characterize the early stages of high-mass star formation, we have observed G023.477+0.114 as part of the ALMA Survey of 70 $\mu$m Dark High-mass Clumps in Early Stages (ASHES). 
We have conducted $\sim$1$\farcs2$ resolution observations with the Atacama Large Millimeter/submillimeter Array (ALMA) at 1.3 mm in dust continuum and molecular line emission. 
We identified 11 cores, whose masses range from 1.1\,$M_\odot$ to 19.0\,$M_\odot$. Ignoring magnetic fields, the virial parameters of the cores are below unity, implying that the cores are gravitationally bound. However, when magnetic fields are included, 
the prestellar cores are close to virial equilibrium, while the protostellar cores remain sub-virialized. 
Star formation activity has already started in this clump. 
Four collimated outflows are detected in CO and SiO. 
H$_2$CO and CH$_3$OH emission coincide with 
the high-velocity components seen in the CO and SiO emission. 
The outflows are randomly oriented for 
the natal filament and the magnetic field. The position-velocity diagrams suggest that episodic mass ejection has already begun even in this  
very early phase of protostellar formation. 
The masses of the identified cores are comparable to the expected maximum stellar mass  that this IRDC could form (8--19\,$M_\odot$). 
We explore two possibilities on how IRDC G023.477+0.114 could eventually form 
high-mass stars in the context of theoretical scenarios.
\end{abstract}

\section{Introduction} \label{sec:intro}

High-mass star formation, especially in the early phases, still remains unclear.
Some theoretical mechanisms aim to explain the formation of high-mass stars. For instance, 
the turbulent core accretion scenario \citep{McKeeTan03} suggests that virialized prestellar high-mass ($\ga30\,M_\odot$) cores, supported by turbulence and/or magnetic fields form high-mass stars. 
On the other hand, the competitive accretion  scenario \citep{Bonnell01} predicts that initially low-mass ($\sim1\,M_\odot$) stellar seeds, which are produced near the bottom of the global gravitational potential of a parent clump, grow into high-mass stars by preferentially acquiring material from the surrounding environment. 

These theoretical scenarios predict distinguishable initial conditions for high-mass star formation (e.g., initial core masses). However, we do not have enough knowledge of the early stages of high-mass star formation from observations. 
Thus, the formation scenario still remains under debate.
Some infrared dark clouds (IRDCs) are thought to be dense quiescent regions prior to active star formation, and suitable to the study of the early stages of high-mass star formation  \citep[][]{Rathborne06, BerginTafalla07}. 
Recent high-angular resolution observations have revealed the  properties of cores embedded in IRDCs  with Submilimeter array (SMA) \citep[][]{Zhang09,Zhang11,Zhang14, Wang11, Lu15, Sanhueza17, Pillai19, Li19a},
with the Combined Array for Research in Millimeter-wave Astronomy (CARMA) \citep[][]{Pillai11, Sanhueza13}, 
and with ALMA \citep[][]{Sakai13, Yanagida14, Zhang15, Svoboda19, Sanhueza19, Rebolledo20, Li21, Redaelli21, Zhang21, Olguin21}. 

To understand the very early phases of high-mass star formation, we have conducted the ALMA Survey of 70 $\mu$m dark High-mass clumps in Early Stages (ASHES). The motivation and the properties of pilot survey are described in \citet{Sanhueza19}. They reported that about half of the cores detected in 12 IRDCs have masses lower than 1 $M_\odot$, and there were no massive ($>$30 $M_\odot$) prestellar cores.  Such observational results favor models in which high-mass stars are formed from low-mass cores (e.g., competitive accretion scenario).
Many outflows are detected even in such 3.6--70 $\mu$m dark IRDCs \citep[e.g.,][]{Li20, Tafoya21}.
As outflows are thought to be accretion-driven, these outflows would enable us to understand the early phase's accretion history which is otherwise extremely difficult to assess, except for a few examples \citep{Contreras18, Liu18}.
The richness of the data allows detailed studies on interesting targets that stand out from the sample. In this paper, we will report a case study one of the 70 $\mu$m dark IRDCs from ASHES, G023.477+0.114 (hereafter G023.477) also known as IRDC 18310-4 with many molecular lines detected, in addition to the dust continuum emission.

G023.477 has been regarded as a prestellar, high-mass clump candidate \citep{Beuther13,Beuther15}. 
Distance estimates for G023.477 disagree.  \citet{Ragan12a} estimate a distance of $4.9\pm0.3$\,kpc, while \citet{Urquhart18} estimate a distance of $5.6\pm0.3$\,kpc.
\citet{Ragan12a} estimated the distance following \citet{Reid09} with a systemic velocity of a  $v_\mathrm{LSR}=86.5\mathrm{\,km\,s^{-1}}$\citep{Sridharan05}, 
and \citet{Urquhart18} used the rotation curve of \citet{Reid14} with a $v_\mathrm{LSR}=85.4\mathrm{\,km\,s^{-1}}$\citep{Wienen12}.
The former $v_\mathrm{LSR}=86.5\mathrm{\,km\,s^{-1}}$ is in agreement with our observations. 
We recalculated the distance using a $v_\mathrm{LSR}=86.5\mathrm{\,km\,s^{-1}}$ and the python-based ``Kinematic Distance Calculation Tool" of \citet{Wenger18}, which evaluates a Monte Carlo kinematic distance adopting the solar Galactocentric distance of 8.31 $\pm$ 0.16\,kpc \citep[][]{Reid14}. 
The estimated near kinematic distance is 5.2 $\pm$ 0.5\,kpc, mostly consistent with reference values.
Considering most studies in G023.477 adopted 4.9\,kpc as the kinetic distance \citep[][]{Ragan12a,Beuther13,Tackenberg14,Beuther15,Beuther18}, we adopt a distance of 4.9\,kpc,  corresponding to a galactocentric distance of $R_\mathrm{GC}=4.3$\,kpc. 

\begin{figure*}
    \epsscale{1.2}
    \plotone{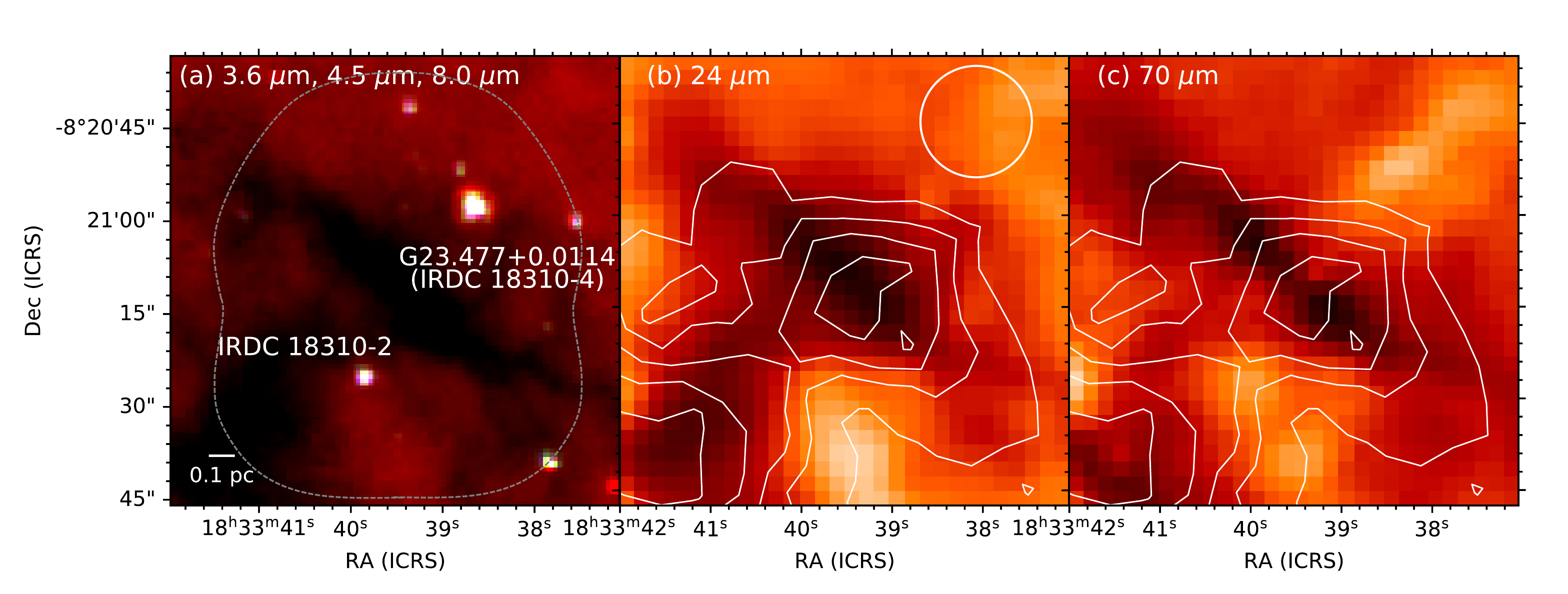}
    \caption{$Spitzer$ and $Herschel$ infrared images for G023.477. 
    (a) $Spitzer$/IRAC three-color (3.6 $\mu$m in blue, 4.5 $\mu$m in green, and 8.0 $\mu$m in red) image. 
    Dashed gray contour represents the area mosaicked with ALMA.
    (b) $Spitzer$/MIPS 24 $\mu$m image. The white contours  are 870 $\mu$m dust continuum emission from the ATLASGAL survey.
    Contour levels for the 870 $\mu$m dust continuum emission are 3, 5, 7, 9 and 12$\sigma$ with 1$\sigma$ = 86.1 mJy\,beam$^{-1}$.
    A white dashed circle on the top right shows the beam size ($\sim$18$\farcs2$) of ATLASGAL survey.
    (c) $Herschel$/PACS 70 $\mu$m image. 
    The white contours are same as those in (b).
    }
    \label{fig:infrared}
\end{figure*}

The region is dark even at 100 $\mu$m wavelength \citep[see Figure~2 in][]{Beuther15} and has a mass of $M_\mathrm{clump}\sim$1000 $M_\odot$\citep{Sridharan05,Yuan17}. 
Figure \ref{fig:infrared} shows the $Spitzer$ and $Herschel$ images of G023.477.
The left panel shows the three color composite diagram (3.6, 4.5, and 8\,$\mu$m) taken in GLIMPSE survey \citep[][]{Benjamin03-GLIMPSE}. 
For a comparison, the center and right panels display the 24 and 70\,$\mu$m emission taken in MIPSGAL \citep[][]{Carey09} and Hi-GAL \citep[][]{Molinari10} survey
, respectively, with contours of 870\,$\mu$m continuum emission obtained by the ATLASGAL survey \citep[][]{Schuller09}.  
The infrared dark region extends from the  north-east to the south-west direction as a filamentary structure. 
In the south-east relative to the center of G023.477, another dense compact clump IRDC 18310-2 is located. 
These two clumps are connected by a 24 $\mu$m dark region. 
The 870\,$\mu$m dust continuum emission also shows elongated structure north-east to south-west.

Within G023.477, at least four cores 
are detected with masses ranging from 9.6 to 19 $M_\odot$ \citep{Beuther13,Beuther15}, after scaling down their gas-to-dust mass ratio of 186 to the typical of 100.
\citet{Beuther15} mentioned that the dense core named mm2, located in north-west from the clump center, has the potential of hosting a protostar because it is slightly brighter in 70 $\mu$m than its surrounding.  
However, since its bolometric luminosity is only about 16 $L_\odot$, the compact and efficient accretion has not begun yet \citep[][]{Beuther15}. 
While Mopra  observations show no sign of outflows \citep{Tackenberg14}, the multiple components of N$_2$H$^+$ (2--1) detected from each core and a virial analysis suggest that the clump is dynamically collapsing and the cores embedded in the clump are in the collapse phase \citep{Beuther13, Beuther15}.  
Additionally, \citet{Beuther18} detected polarized emission from all the four cores in this region, suggesting that the magnetic field plays a role in the fragmentation and collapse process.  
The narrow linewidths of N$_2$H$^+$ (3--2) \citep{Beuther15} also suggest that turbulence plays a minor role in supporting the cores against gravitational collapse.  

In this paper, we reveal the detailed structure of G023.477 using ALMA Band 6 (1.3\,mm) observations of dust emission, deuterated molecular lines, and outflow tracers. 
We describe the observations in Section~\ref{sec:obs} and show the results in Section~\ref{sec:results}.
In Section~\ref{sec:densecores}, we identify dust cores from 1.3 mm continuum emission and estimate physical parameters using continuum emission, DCO$^+$, H$_2$CO, and C$^{18}$O. 
We also discuss the distribution of the deuterated molecules.
The detection of outflows is presented in Section~\ref{sec:outflow}.
In Section~\ref{sec:position}, we investigate the orientation of the outflows compared with the position angles of the filament and the magnetic fields.
We also describe the evolutionary stages of cores in G023.477 in Section~\ref{sec:evolution}, and discuss the potential for 
high-mass star formation in Section~\ref{sec:highmass}. Section~\ref{sec:summary} presents a summary of our work. 

\section{Observations}
\label{sec:obs}

\begin{deluxetable*}{lcc}
\label{tab:obspara}
\tabletypesize{\footnotesize}
\tablecaption{ Observing Parameters}
\tablewidth{0pt}
\tablehead{
\colhead{Parameters} & \colhead{ACA} & \colhead{ALMA 12\,m array}}
\startdata
Observing date (YYYY-MM-DD)		&	2018-10-22 / 23-24 & 2019-03-12 \\
Number of antennas				&	11/10         	& 45 		\\
Primary beam size (arcsec)      &	44\farcs6	        &  25\farcs2	\\
Bandpass calibrators  		    & J1924-2914	& J1751+0939 \\
Flux and Phase calibrators    	& J1911-2006       & J1743-0350	\\
Baselines (m) & 8.9--48.9/8.9--45.0 & 15.0--313.7\\
Total on-source time (minutes)      	& 29    & 13
\enddata
\end{deluxetable*}

We have used the ASHES survey data from the Cycle 6 project (2018.1.00192.S, PI:\,P. Sanhueza).  
The band 6 (1.3\,mm) observations were made on 2019 March 12 (ALMA 12\,m array), 2018 October 22 to 24 (Atacama Compact 7 m array, hereafter the ACA), and  2018 October 30 (total power, TP).   
The phase reference center for the mosaic is R.A. (J2000.0) = $18^h 33^m 39 \fs 532$ and Dec (J2000.0) = $-08^\circ 21^\prime 09\farcs 60$. The observing parameters are listed in Table~\ref{tab:obspara}. 

The whole IRDC was covered by a 10-pointing and 3-pointing mosaics with the ALMA 12\,m array and ACA, respectively.
The ALMA 12\,m array consisted of 45 antennas, with a baseline ranging from 15 to 313 m. The flux calibration and phase calibration were carried out using J1743-0350.
The quasar J1751+0939 was used for bandpass calibration. The total on source time was $\sim$13 minutes. More extended continuum and line emission were recovered by including the ACA data.  
The 7 m array observations consisted of 10 or 11 antennas, with baselines ranging from 9 to 49 m. The flux calibration and phase calibration were carried out using J1911-2006, and the bandpass calibration was carried out using J1924-2914.
The total on source time was $\sim$29 minutes for ACA. 
These observations are sensitive to angular scales smaller than $\sim$11$''$ and $\sim$19$''$, respectively.

Our spectral setup includes 13 different molecular lines: $^{13}$CS ($J$\,=\,5--4), N$_2$D$^+$($J$\,=\,3--2), CO ($J$\,=\,2--1), DCN ($J$\,=\,3--2), CCD ($N$\,=\,3--2), DCO$^+$($J$\,=\,3--2), SiO ($J$\,=\,5--4), H$_2$CO ($J_\mathrm{K_a,K_c}$=\,3$_{0,3}$--$2_{0,2}$), H$_2$CO ($J_\mathrm{K_a,K_c}$=\,3$_{2,2}$--$2_{2,1}$), H$_2$CO ($J_\mathrm{K_a,K_c}$=\,3$_{2,1}$--$2_{2,0}$), CH$_3$OH ($J_\mathrm{K}$=\,4$_{2}$--$3_{1}$), HC$_3$N ($J$=24--23), and C$^{18}$O ($J$\,=\,2--1). 
We summarize the spectral window setting in Table \ref{tab:spw}.
The velocity resolution of CO, C$^{18}$O, CH$_3$OH, H$_2$CO, and HC$_3$N is $\sim$1.3 km\,s$^{-1}$, that of $^{13}$CS and N$_2$D$^+$ is 0.079 km\,s$^{-1}$, and that of other molecules is $\sim$0.17 km\,s$^{-1}$.

Data reduction was performed using the CASA software package versions 5.4.0 for calibration and 5.6.0 for imaging \citep{McMullin07}.
The continuum image was obtained by averaging line-free channels with a Briggs\textquotesingle s robust weighting of 0.5 to the visibilities. 
The effective bandwidth for continuum emission was 3.64\,GHz.
An average 1$\sigma_\mathrm{cont}$ root mean square (rms) noise level of
0.093 mJy beam$^{-1}$ was achieved in the combined 12 and 7\,m array continuum image. 
The synthesized beam size is $1\farcs 4 \times 1\farcs 1$ with a position angle (P.A.) of $\sim$77\arcdeg, with a geometric mean of  $1\farcs 2$ that corresponds to  $\sim$5900 au in linear scale at the source distance. 
For molecular lines, we used the automatic cleaning algorithm for imaging data cubes, YCLEAN \citep[][]{Contreas_yclean_18, Contreras18} to CLEAN the data cubes for each spectral window with custom made masks. 
We adopted a Briggs\textquotesingle s robust weighting of 2.0 (natural weighting) to improve the S/N ratio.
The channel widths used for measuring the noise level are $\sim$0.66\,km\,s$^{-1}$ for CO,  C$^{18}$O, HC$_3$N, H$_2$CO and CH$_3$OH, and $\sim$0.17\,km\,s$^{-1}$ for the other lines, 
resulting in an average 1$\sigma$ rms noise level of 3.8 mJy beam$^{-1}$ and 7.0 mJy beam$^{-1}$, respectively. 
The velocity resolution is two times coarser than the channel width due to a Hanning filter applied by ALMA observatory (ALMA science primer\footnote{https://almascience.nao.ac.jp/documents-and-tools/cycle6/alma-science-primer}), but we smoothed the cubes of deuterated molecules to boost the S/N ratio.
The average synthesized beam size is $1\farcs 6 \times 1\farcs 2$ (P.A. $\sim$67\arcdeg). 
The rms noise level ($\sigma$) measured in the line-free channels for each line and the beam size of each spectral windows are also summarized in Table~\ref{tab:spw}.
All images shown in the paper are the ALMA 12\,m and ACA combined, prior to the primary beam correction, while all measured fluxes are derived from the combined data and corrected for the primary beam attenuation.

\begin{deluxetable*}{lcccccc}
\label{tab:spw}
\tabletypesize{\footnotesize}
\tablecaption{Summary of spectral windows}
\tablewidth{0pt}
\tablehead{
\colhead{Transition}  & \colhead{Rest Frequency} & \colhead{Bandwidth} & \colhead{Velocity Resolution} & \colhead{$E_{\mathrm{u}}/k$} & \colhead{RMS Noise Level ($\sigma$)} & \colhead{Beam Size} \\ 
\colhead{} & \colhead{GHz} & \colhead{GHz} & \colhead{\,km\,s$^{-1}$} & \colhead{K} & \colhead{mJy\,beam$^{-1}$} & \colhead{arcsec$\times$arcsec}} 
\startdata
DCO$^+$ ($J$\,=\,3--2) & 216.112580 & 0.059 & 0.169 & 20.74 & 6.89 & 1.66$\times$1.26\\
CCD ($N$\,=\,3--2) & 216.373320 & 0.059 & 0.169 & 20.77 & 6.88 & 1.66$\times$1.26\\
SiO ($J$\,=\,5--4) & 217.104980 & 0.059 & 0.169 & 31.26 & 5.79 & 1.65$\times$1.26\\
DCN ($J$\,=\,3--2) & 217.238530 & 0.059 & 0.168 & 20.85 & 6.38 & 1.65$\times$1.25\\
H$_2$CO ($J_\mathrm{K_a,K_c}$=\,3$_{0,3}$--$2_{0,2}$) & 218.222192 & 1.875 & 1.338 & 20.96 & 2.76 & 1.65$\times$1.26\\
HC$_3$N ($J$\,=\,24--23) & 218.324720 & 1.875 & 1.338 & 130.98 & 2.76 & 1.65$\times$1.26\\
CH$_3$OH ($J_\mathrm K$=\,4$_2$--$3_1$) & 218.440063 & 1.875 & 1.338 & 45.46 & 2.76 & 1.65$\times$1.26 \\
H$_2$CO ($J_\mathrm{K_a,K_c}$=\,3$_{2,2}$--$2_{2,1}$) & 218.475632 & 1.875 & 1.338 & 68.09 & 2.76 & 1.65$\times$1.26\\
H$_2$CO ($J_\mathrm{K_a,K_c}$=\,3$_{2,1}$--$2_{2,0}$) & 218.760066 & 1.875 & 1.338 & 68.11 & 2.76 & 1.65$\times$1.26\\
C$^{18}$O ($J$\,=\,2--1) & 219.560358 & 1.875 & 1.338 & 15.81 & 3.73 & 1.64$\times$1.25\\
CO ($J$\,=\,2--1)  &230.538000  & 1.875 & 1.268 & 16.60 & 2.64 & 1.55$\times$1.20\\
$^{13}$CS ($J$\,=\,5--4) & 231.220686 & 0.059 & 0.079 & 33.29 & 6.62 & 1.55$\times$1.19 \\
N$_2$D$^+$ ($J$\,=\,3--2)  & 231.321828 & 0.059 & 0.079 & 22.20 & 8.09 &  1.56$\times$1.19
\enddata
\end{deluxetable*}

\section{Spatial Distribution} \label{sec:results}
\subsection{Dust continuum emission}

\begin{figure*}
\epsscale{1.0}
 \plotone{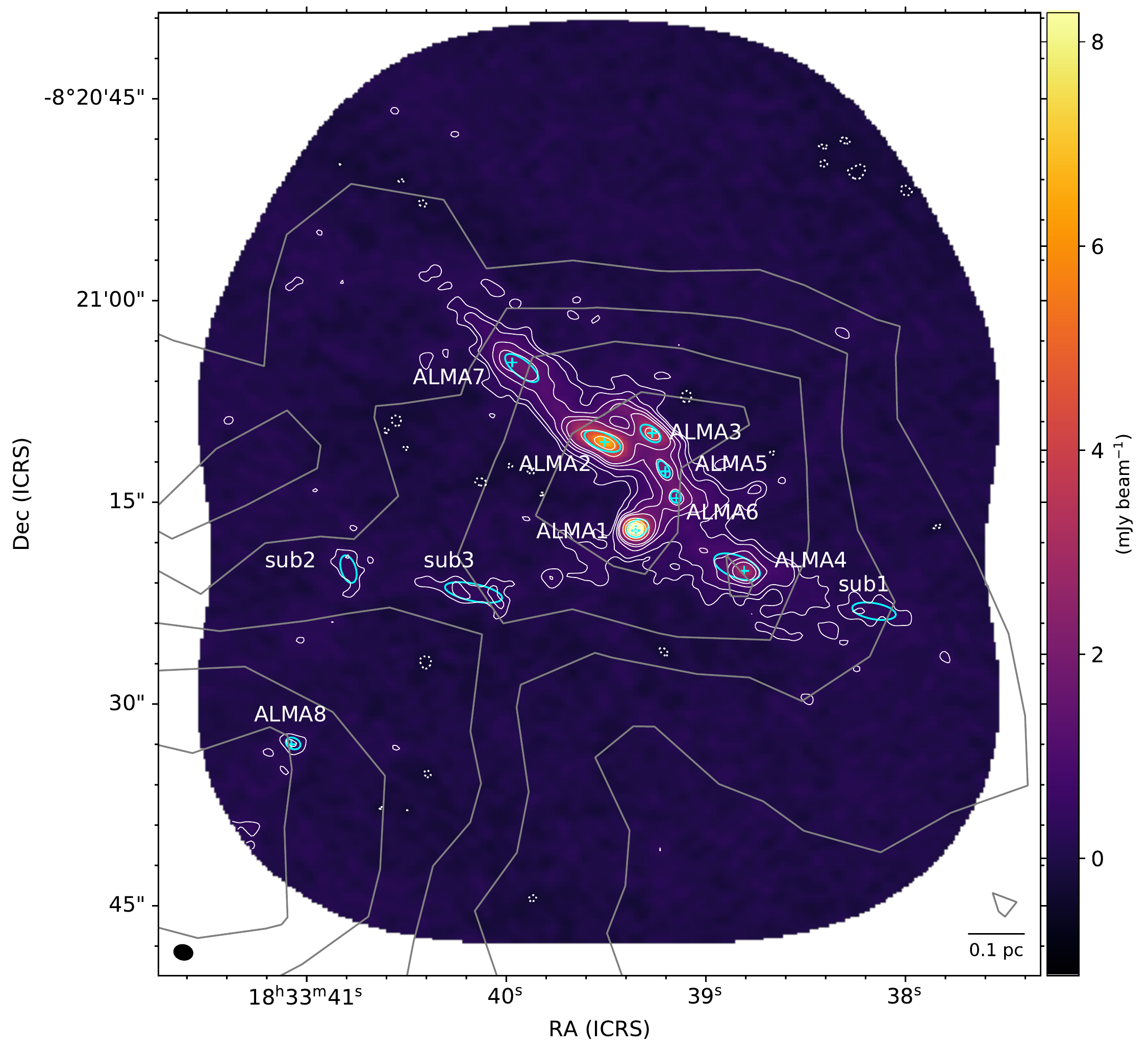}
    \caption{ALMA 1.3 mm continuum image in white contours (-3, 3, 5, 10, 15, 20, 40, 60, 80, 100 and 120$\sigma_\mathrm{cont}$} with 1$\sigma_\mathrm{cont}$ = 0.093 mJy beam$^{-1}$). 
    The dotted contours show the negative components.
     The cyan ellipses represent the identified cores by dendrogram algorithms (Section~\ref{sec:identification}), and 
    the plus symbols show the continuum peak position of ALMA1--8.
    The gray contours show the 870\,$\mu$m continuum emission from the ATLASGAL survey, and contour levels are the same as in Figure\,\ref{fig:infrared}. 
    The black ellipse in the bottom left corner represents the synthesized beam size.
    The spatial scale is indicated by the black line in the bottom right corner. 
 \label{fig:continuum} 
\end{figure*}

Figure~\ref{fig:continuum} presents the ALMA 1.3 mm continuum image.  
This region has a prominent filamentary structure (hereafter main filament) running from the north-east to the south-west direction and a chain of faint condensed structures 
in the east-west direction that connects to the main filament near its center. 
This kind of structure is roughly consistent with the large-scale dust emission observed with the single-dish APEX telescope at 870 $\mu$m (ATLASGAL) and with the infrared dark region (Figure~\ref{fig:infrared}). 
The chain elongated in the east-west direction corresponds to the bridge of two IRDCs as mentioned in the introduction.
At the intersection, the dust continuum emission takes its maximum at 12\,mJy\,beam$^{-1}$. 
Our mosaicked observations revealed a whole picture of G023.477 with a wide field of view. Our high-angular resolution observations unveiled several compact substructures  embedded in the filamentary IRDC that will likely form stars, i.e., dense cores. 
In Section~\ref{sec:densecores}, we identify cores using the dendrogram technique. 

\subsection{Molecular line emission}\label{sec:line_emission}
\begin{figure*}
    \epsscale{1.1}
    \plotone{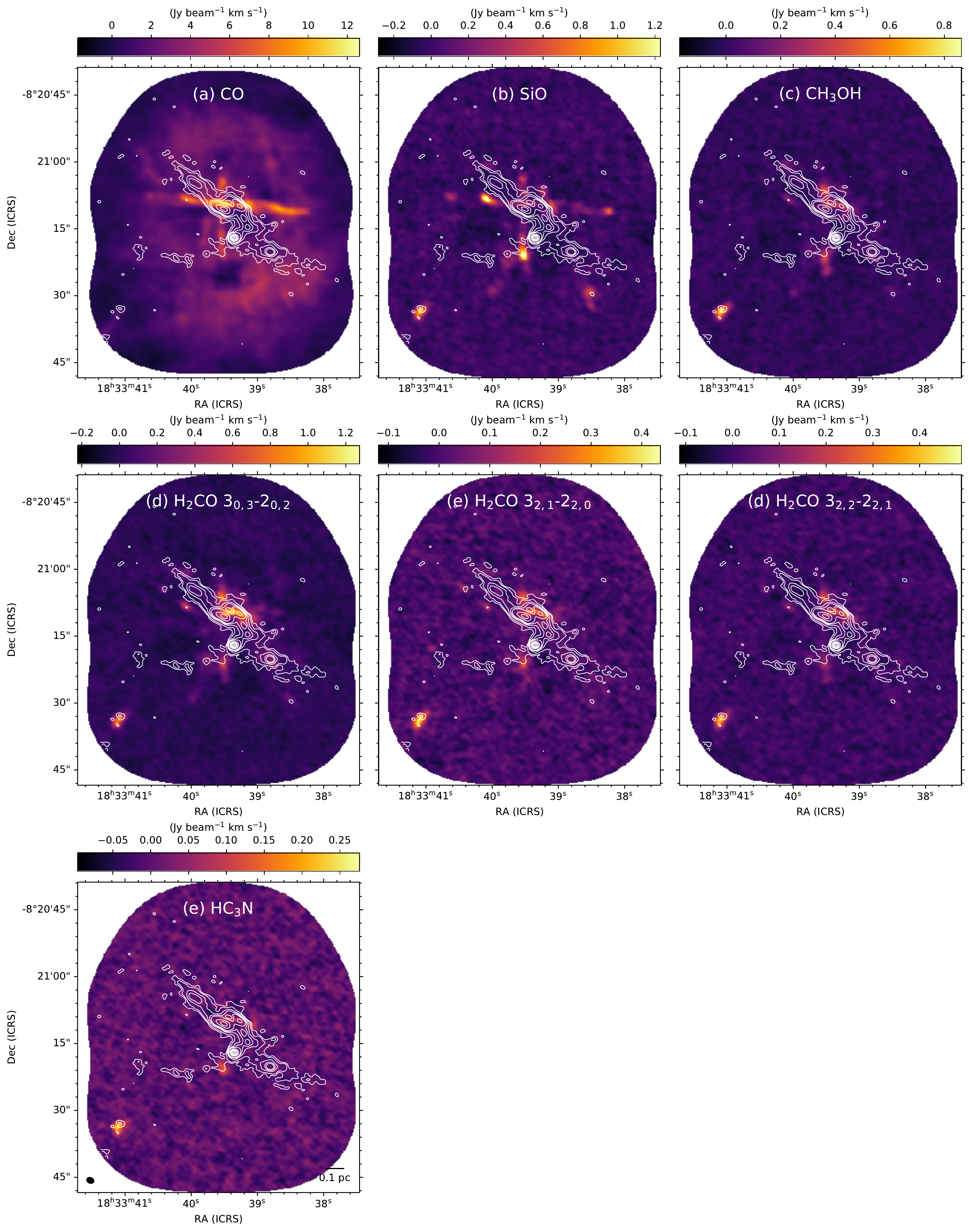}
    \caption{Integrated intensity maps of (a) CO ($J$=2--1), (b) SiO ($J$=5--4), (c) CH$_3$OH ($4_2$--$3_1$), (d) H$_2$CO ($3_{0,3}$--$2_{0,2}$), (e) H$_2$CO ($3_{2,1}$--$2_{2,0}$), (f) H$_2$CO ($3_{2,2}$--$2_{2,1}$), and (g) HC$_3$N ($J$=24--23).
    The integrated velocity ranges are 20\,km\,s$^{-1} < v_\mathrm{LSR}<$181\,km\,s$^{-1}$ for CO,  47\,km\,s$^{-1} < v_\mathrm{LSR}<$126\,km\,s$^{-1}$ for SiO, and $|v_\mathrm{LSR}-v_\mathrm{sys}|\la$10\,km\,s$^{-1}$ for H$_2$CO, CH$_3$OH, and HC$_3$N, where $v_\mathrm{sys}$ is the systemic velocity of this region of 86.5 $\mathrm{ km\,s^{-1}}$.
    The white contours show the 1.3 mm continuum emission and the levels are the same as those in Figure \ref{fig:continuum}.
    The synthesized beam size and the spatial scale are shown in the lower left panel.}
    \label{fig:outflow_tracers}
\end{figure*}
Figure \ref{fig:outflow_tracers} shows the integrated intensity maps 
of CO ($J$=2--1), SiO ($J$=5--4), CH$_3$OH ($J_\mathrm{K}$=\,$4_2$--$3_1$), H$_2$CO ($J_\mathrm{K_a,K_c}$=\,$3_{0,3}$--$2_{0,2}$), H$_2$CO ($J_\mathrm{K_a,K_c}$=\,$3_{2,1}$--$2_{2,0}$), H$_2$CO ($J_\mathrm{K_a,K_c}$=\,$3_{2,2}$--$2_{2,1}$), and HC$_3$N ($J$=24--23) which are often used as molecular outflow tracers  \citep[e.g.,][]{Tafalla10,Sanhueza10,Zhang15,Cosentino18,Tychoniec19, Li19a, Li20}. 
For each line, we integrated the emission greater than 4$\sigma$ in the following velocity ranges, where $\sigma$ is the rms noise level in the line-free channels (Table~\ref{tab:spw}). We determined this threshold by checking the cubes to avoid noise contamination. One example of the channel map is Figure~\ref{fig:channelC18O} in Appendix, from which we detemined the integration range.
The integrated velocity ranges are 20\,km\,s$^{-1} < v_\mathrm{LSR} <$181\,km\,s$^{-1}$ for CO, and 
47\,km\,s$^{-1} < v_\mathrm{LSR} <$126\,km\,s$^{-1}$ for SiO.
As for H$_2$CO, CH$_3$OH, and HC$_3$N, we integrated the emission in the range of $|v_\mathrm{LSR}-v_\mathrm{sys}|\la$10\,km\,s$^{-1}$, where $v_\mathrm{sys}$ is the systemic velocity of this region of 86.5 $\mathrm{ km\,s^{-1}}$\citep{Sridharan05}.

Two collimated structures in the north-south and east-west direction are easily detected in CO emission, as shown in Figure~\ref{fig:outflow_tracers} (a). 
SiO emission is also found along such linear structures and especially trace the regions where CO emission is strongly detected. 
The maximum velocity of CO and SiO emission with respect to the systemic velocity ($|v_\mathrm{LSR}-v_\mathrm{sys}|$) is over 90\,km\,s$^{-1}$ and 40\,km\,s$^{-1}$, respectively (see Appendix for additional details). This high velocity gas is likely gravitationally unbound, implying outflows or jets. We identify outflows in Section~\ref{sec:outflow}. 
The CH$_3$OH and H$_2$CO emission are also bright in north-south direction and in the crossing point of the two collimated structures as traced in CO and SiO.

\begin{figure*}
    \epsscale{1}
    \plotone{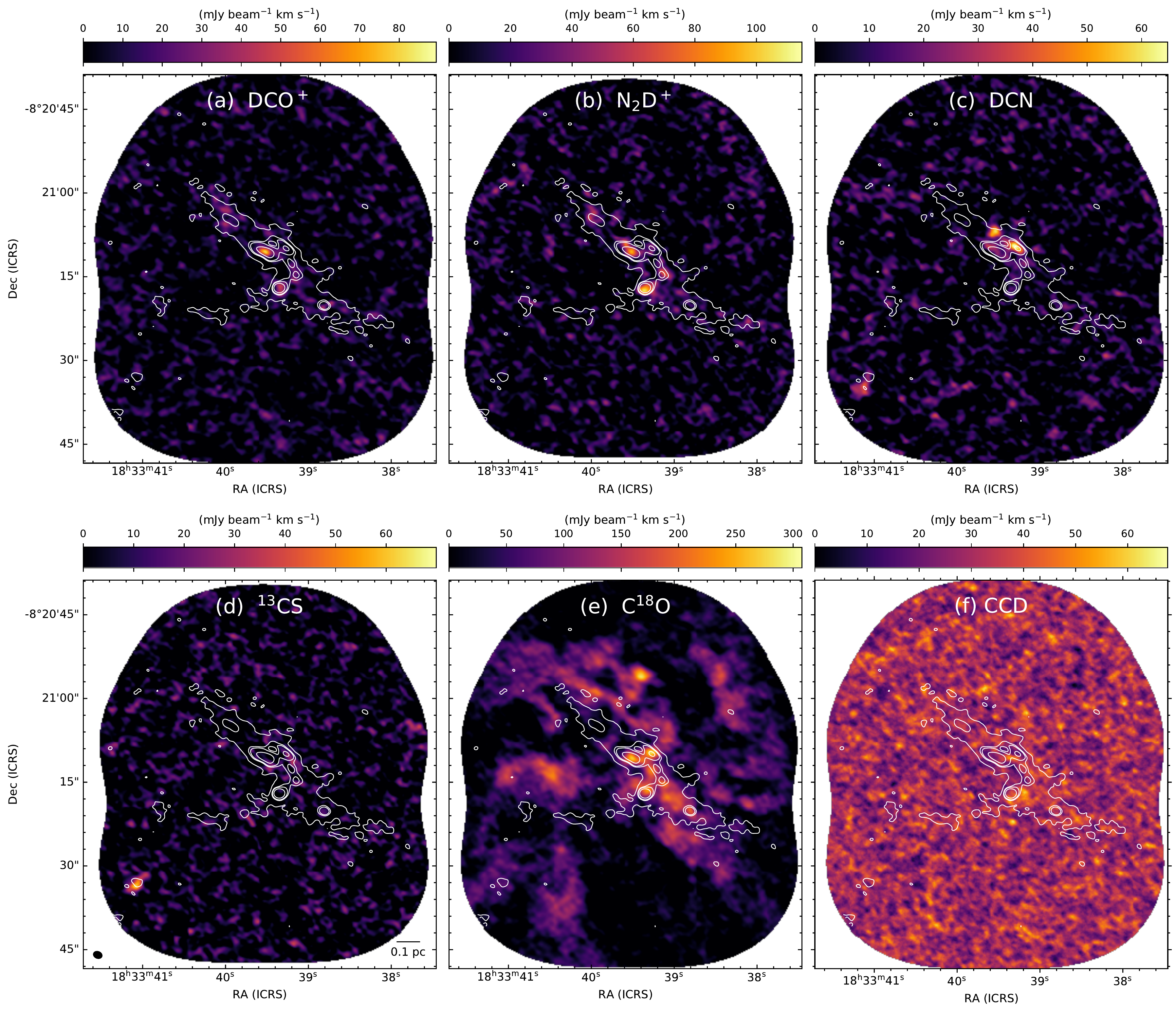}
    \caption{Integrated intensity maps of (a)\,DCO$^+$ ($J$=3--2), (b)\,N$_2$D$^+$ ($J$=3--2), (c)\,DCN ($J$=3--2), (d)\,$^{13}$CS ($J$=5--4), (e)\,C$^{18}$O ($J$=2--1), and (f) CCD ($N$=3--2)}. The integrated velocity ranges are 84.2\,km\,s$^{-1} < v_\mathrm{LSR} < 90.4$\,km\,s$^{-1}$ for N$_2$D$^+$, DCO$^+$, DCN, $^{13}$CS, and CCD, and 82\,km\,s$^{-1} < v_\mathrm{LSR} < 92$\,km\,s$^{-1}$ for C$^{18}$O. The white contours show the 1.3 mm continuum emission and the levels are 3, 15, 20, and 40$\sigma_\mathrm{cont}$ with 1$\sigma_\mathrm{cont}$ = 0.093 mJy beam$^{-1}$.
    The spatial scale and the beam size are shown at the bottom in the left bottom panel.
    \label{fig:densegas_mom0}
\end{figure*}
Figure \ref{fig:densegas_mom0} shows the integrated intensity maps of N$_2$D$^+$ ($J$=3--2), DCO$^+$ ($J$=3--2), DCN ($J$=3--2), $^{13}$CS ($J$=5--4), C$^{18}$O ($J$=2--1), and CCD ($N$=3--2) overlaid with contours of the 1.3 mm continuum emission presented in Figure\,\ref{fig:continuum}, which, except for C$^{18}$O, are used as dense gas tracers due to their high critical densities.  
The integrated velocity ranges are 84.2\,km\,s$^{-1} < v_\mathrm{LSR} < 90.4$\,km\,s$^{-1}$ for N$_2$D$^+$, DCO$^+$, DCN, $^{13}$CS, and CCD, and 82\,km\,s$^{-1} < v_\mathrm{LSR} < 91$\,km\,s$^{-1}$ for C$^{18}$O, where the emission is greater than 4$\sigma$. 
The peak intensities are weaker than lines in Figure \ref{fig:outflow_tracers}.
The spatial distributions of N$_2$D$^+$, DCO$^+$, DCN, and $^{13}$CS are compact, and agree well with dust continuum emission, while C$^{18}$O is more extended.
The local peaks of DCO$^+$, DCN, and N$_2$D$^+$ emission coincide with the dust continuum peaks. 
In particular, the N$_2$D$^+$ peak emission lies 
at the intersection between the main filament and the chain of condensed structure.
There is no significant $^{13}$CS emission  associated with 
the main filament. 
On the other hand, relatively strong and compact $^{13}$CS emission is detected around the continuum emission located near the south-east of the observed area. 
The C$^{18}$O emission is distributed throughout the entire region, having  both compact and extended components, although the emission does not follow the main filament well.  
Specifically, the emission is weak at the northern part of the main filament. 
Multiple velocity components along the line of sight are found 
(see the channel maps presented in Appendix).
There is no CCD emission higher than 3$\sigma$ in the field of view. 

\section{Dense cores} \label{sec:densecores}
\subsection{Core Identification} \label{sec:identification}
To define the dust cores, we adopt the dendrogram technique \citep{Rosolowsky08}.
There are three main parameters, $F_\mathrm{min}$, $\delta$, and $S_\mathrm{min}$. $F_\mathrm{min}$ sets the minimum value above which we define structures 
and $\delta$ sets a minimum significance to separate them.  
$S_\mathrm{min}$ is the minimum number of pixels to be contained in the  smallest individual structure (defined as leaf in dendrogram).  
Given the influence of the noise, the minimum acceptable significance should be at least of 2 signal-to-noise ratios \citep{Rosolowsky08}. 
We adopt 3$\sigma_\mathrm{cont}$ for $F_\mathrm{min}$, 2$\sigma_\mathrm{cont}$  for $\delta$ (with 1$\sigma_\mathrm{cont}$ = 0.093 mJy beam$^{-1}$), and the number of pixels contained in half of the synthesized beam for $S_\mathrm{min}$. 
The smallest structures identified in the dendrogram, leaves, are defined as cores, corresponding to cyan ellipses in Figure~\ref{fig:continuum}.

With the conditions mentioned above, we identify eleven cores 
(all with flux densities above 3.5$\sigma_\mathrm{cont}$). The cores with the peak intensity higher than 10$\sigma_\mathrm{cont}$ are named ALMA1$-$8, while the remaining ones are named sub1-3. 
ALMA1, ALMA2, ALMA3, and ALMA7 correspond to mm3, 1, 2, and 4 in \cite{Beuther13}, respectively, and ALMA4 is identified as mm4 in \cite{Beuther18}. 
If we set the synthesized beam size for $S_\mathrm{min}$ without changing the other two dendrogram parameters, only ALMA6 would be excluded.
Hereafter, we will mainly discuss ALMA1-8. 
In Table~\ref{tab:dendro}, we summarize the continuum peak position, peak intensity, flux density, deconvolved sizes, and the position angles, which are measured by the dendrogram algorithm, in addition to the  corresponding source names reported in \citet{Beuther18}. 
The deconvolved size is computed from the intensity weighted second moment in direction of greatest elongation in the PP plane (major axis) and perpendicular to the major axis (minor axis), see additional details in the astrodendro website.\footnote{https://dendrograms.readthedocs.io/en/stable/}

The integrated intensity of the combined data sets (12 m + ACA) over the region 
is 1.2 times larger than the 12 m only image. 
We estimated how much flux is recovered by ALMA by comparing the 1.2\,mm integrated intensity ($F_\mathrm{1.2\,mm}$) obtained with IRAM 30\,m telescope \citep[][]{Beuther02} assuming a dust emissivity spectral index ($\beta$) of 1.5 as $F_\mathrm{1.3\,mm, ALMA}$/$F_\mathrm{1.3\,mm, exp}$, where $F_\mathrm{1.3\,mm, ALMA}$ is the observed 1.3 mm integrated intensity obtained by ALMA and  $F_\mathrm{1.3\,mm, exp}$ is estimated as $F_\mathrm{1.3\,mm, exp}$=$F_\mathrm{1.2\,mm}$(1.3/1.2)$^{-1.5}$. 
The flux recovered by ALMA is 31$\%$. Comparing with the ATLASGAL 870 $\mu$m emission, the recovered flux is 27$\%$, consistent with SMA/ALMA observations in other IRDC studies \citep[e.g.,][]{Sanhueza17, Liu18,Sanhueza19}. 

\begin{deluxetable*}{lccccccc}
\tabletypesize{\footnotesize}
\tablecaption{ALMA 1.3 mm continuum sources}
\tablewidth{0pt}
\tablehead{
 & \colhead{R.A.}  & \colhead{Decl.} & \colhead{Peak Intensity} & \colhead{Flux density}& \colhead{Deconvolved Size} & \colhead{Position Angle}  & \colhead{Other Source Names \tablenotemark{a}}\\ 
& J2000.0 & J2000.0 & mJy beam$^{-1}$ & mJy & arcsec $\times$ arcsec & deg & }
\startdata
ALMA1 & 18 33 39.53 & -08 21 17.10 & 12 & 16 & 1.4 $\times 1.0 $ & -150 & mm3 \\
ALMA2 & 18 33 39.51 & -08 21 10.51 & 6.7 & 21 &  2.4 $\times 1.1 $ & 160 & mm1 \\
ALMA3 & 18 33 39.27 & -08 21 09.85 & 4.5 & 7.6 &  1.5 $\times 0.83 $ & 140 & mm2 \\
ALMA4 & 18 33 38.81 & -08 21 20.10 & 3.7 & 12 & 3.1 $\times 1.4 $ & 160 & mm4   \\
ALMA5 & 18 33 39.20 & -08 21 12.70 & 2.2 & 2.6 &  1.2 $\times 0.59 $ & 120 &  \\
ALMA6 & 18 33 39.14 & -08 21 14.60 & 2.2 & 1.9 &  0.79 $\times 0.65 $ & 120 & \\
ALMA7 & 18 33 39.97 & -08 21 04.60 & 1.9 & 7.4 &  2.6 $\times 1.0 $ & 140 &  \\
ALMA8 & 18 33 41.10 & -08 21 33.00 & 1.2 & 1.2 &  0.97 $\times 0.69 $ & 170 & \\
sub1 & 18 33 38.25 & -08 21 23.28 & 0.86 & 3.0 &  2.8 $\times 1.0 $ & 170 & \\
sub2 & 18 33 40.81 & -08 21 19.28 & 0.81 & 1.6 &  1.8 $\times 0.93 $ & 110 &  \\
sub3 & 18 33 40.24 & -08 21 21.88 & 0.65 & 2.6 &  2.7 $\times 1.1 $ & 160 &    \\
\enddata
\tablenotetext{a}{\citet{Beuther18}}
\label{tab:dendro}
\end{deluxetable*}

\subsection{Core physical properties}\label{sec:mass}
Assuming optically thin dust thermal emission and a single dust temperature,
we can estimate the gas mass from the flux density $F_\mathrm{1.3\,mm}$ using
\begin{equation}
M_\mathrm{core} = \mathbb{R}\frac{F_\mathrm{1.3\,mm}d^2}{\kappa_\mathrm{ 1.3\,mm}B_\mathrm{1.3\,mm}(T_\mathrm{dust})},
\end{equation}\\
where $\mathbb{R}$, $d$, $\kappa_\mathrm{1.3mm}$, and  $B_\mathrm{1.3mm}(T_\mathrm{dust})$ are the gas-to-dust mass ratio, the distance to the source \citep[4.9\,kpc,][]{Ragan12a}, 
absorption coefficient of the dust per unit mass, and the Planck function as a function of the dust temperature $T_\mathrm{dust}$, respectively. 
We adopt a gas-to-dust mass ratio, $\mathbb{R}$, of 100 and a dust opacity, $\kappa_\mathrm{1.3mm}$, of 0.9 cm$^2$\,g$^{-1}$ 
from the dust coagulation model of the MRN \citep{Mathis77}
distribution with thin ice mantles at a number density of
10$^6$\,cm$^{-3}$ computed by \citet{Ossenkopf94}.
We conducted SED fitting of HiGAL and ATLASGAL surveys, using $Herschel$ and APEX telescopes, at the peak position of the 870 $\mu$m intensity map. 
The fitting result is Figure~\ref{fig:SED} in Appendix.
The measured fluxes are 646.1 MJy sr$^{-1}$ at 160 $\mu$m, 952.7 MJy sr$^{-1}$ at 250 $\mu$m, 720.2 MJy sr$^{-1}$ at 350 $\mu$m, 340.8 MJy sr$^{-1}$ at 500 $\mu$m, and 60.8 MJy sr$^{-1}$ at 870 $\mu$m.
We determine a dust temperature of $13.8 \pm 0.8$ K at the angular resolution of  35$\farcs0$. The uncertainty is calculated as \citet[]{Guzman15}.

We adopt this temperature to calculate the masses of the identified cores.
The molecular density, $n(\mathrm{H_2})$, was calculated with the assumption that each core is a uniform sphere.
The peak column density, $N_\mathrm{H_2, peak}$, was estimated as
\begin{equation}
    N_\mathrm{H_2, peak} = \mathbb{R}\,\frac{ F_\mathrm{1.3\,mm,\,peak}}{\Omega\, \bar{m}_\mathrm{H_2}\kappa_\mathrm{1.3\,mm}B_\mathrm{1.3\,mm}(T_\mathrm{dust})},
\end{equation}
where $F_\mathrm{1.3\,mm, peak}$ is the peak flux measured at the continuum peak, $\Omega$ is the beam solid angle and $\bar{m}_\mathrm{H_2}$ is the mean molecular mass per hydrogen molecule.
Here, we adopt $\bar{m}_\mathrm{H_2} = 2.8\,m_\mathrm{H}$ \citep[][]{Kauffmann08}.

Core physical parameters are summarized in Table ~\ref{tab:phypara}.
The core radius (R) is defined as half of the geometric mean of the deconvolved size from Table~\ref{tab:dendro}.
The calculated core masses range from 1.1 to 19 $M_\odot$.
Peak column densities are between 0.33 $\times 10^{23}$ and 4.8 $\times 10^{23}$ cm$^{-2}$. 
The number density of the cores ranges from 5.8 $\times$ $10^5$ to 1.7 $\times 10^7$cm$^{-3}$. If we assume 20 K instead of  the computed {\it Herschel} dust temperature of 13.8 K, we obtain masses and number densities 40$\%$ lower.
These core masses and sizes are in agreement with those estimated from cores in other IRDCs \citep[e.g.,][]{Ohashi16,Sanhueza19, Chen19}.

The major sources of uncertainty in the mass calculation come from the gas-to-dust mass ratio and the dust opacity. Assuming that all possible values of $\mathbb{R}$ and $\kappa_\mathrm{1.3mm}$ are distributed uniformly between the extreme values; $70<\mathbb{R}<150$ and $0.7<\kappa_\mathrm{1.3mm}<1.05$ \citep[e.g.,][]{Devereux90,Ossenkopf94, Vuong03}, the standard deviation can be estimated \citep{Sanhueza17}. We adopt the uncertainties derived by \citet{Sanhueza17} of 23\,\% for the gas-to-dust mass ratio and  of 28\,\% for the dust opacity, with respect to the adopted values of 100 and 0.9 cm$^2$\,g$^{-1}$, respectively.
In addition, considering an absolute flux uncertainty of 10\,\% for ALMA observations in band 6, a temperature uncertainty of 6\,\%, and a distance uncertainty of 10\,\%, we estimate a mass and a number density uncertainty of $\sim$50\,\% \citep[see][for more details]{Sanhueza17,Sanhueza19}.

\movetabledown=30mm
\begin{rotatetable*}
\begin{deluxetable*}{lccccccccccccccc}
\label{tab:phypara}
\tabletypesize{\footnotesize}
\tablecaption{Physical Parameters}
\tablewidth{0pt}
\tablehead{
 & \colhead{$M_\mathrm{core}$}& \colhead{$R$} &\colhead{$N_{\mathrm{H_2, peak}}$ \tablenotemark{a}} & \colhead{$n_{\mathrm{H_2}}$} & $\sigma_\mathrm{DCO^+}$ & $\sigma_\mathrm{tot}$ & $v_{\mathrm{core}}$ & $M_\mathrm{k}$ &\colhead{$\alpha_\mathrm{k}$} &\colhead{ $\alpha_{\mathrm{k+B}}$} & \colhead{$N_{\mathrm{H_2CO}}$} &\colhead{$T_{\mathrm{rot}}$} &\colhead{$N_{\mathrm{C^{18}O}}$} & \colhead{$f_\mathrm{C^{18}O}$} & \colhead{Evolutionary\tablenotemark{b}}\\ & $M_\odot$ & 10$^{-2}$ pc & 10$^{23} $cm$^{-2}$ & 10$^6 $cm$^{-3}$ & km s$^{-1}$ & km s$^{-1}$ & km\,s$^{-1}$& $M_\odot$ & & & 10$^{12}$\,cm$^{-2}$ & K & 10$^{14} $cm$^{-2}$ & &Stages}
\startdata
ALMA1 & 14  & 1.4  &   4.8   & 17  &   0.34  &   0.40  &   87.0  &   2.7   &   0.19  &   0.48  &   2.8  &   62   &    8.9    &    300  & (ii)\\
ALMA2 & 19  & 1.9  &   2.7   & 9.2 &   0.63  &   0.66   &   87.4  &   9.8   &   0.53  &   0.83  &   9.1  &   59   &    13    &    110   & (i)\\
ALMA3 & 6.6 & 1.3  &   1.9   & 9.8 &  --  &  --  &  -- & -- &  --  &  --  & 14  &   62   &   15  &   68    & (i)\\
ALMA4 & 11  & 2.4  &   1.5   & 2.5 &   0.45  &   0.50   &    88.2   &   7.1   &   0.66   &    1.8   &   3.8  &   43   &  14   &   59    & (i)\\
ALMA5 & 2.3 & 0.99 &   0.90  & 8.1 &  --  &  --  &  --  & -- &  --  &  --  &   3.6  &   37   &   5.6   &    89    & (ii)\\
ALMA6 & 1.7 & 0.85 &   0.89  & 9.5 &   0.23  &   0.31  &   87.0  &   0.94  &   0.56  &   1.4  & --  & --  &   9.2   &   53    & (iii)\\
ALMA7 & 6.4 & 1.9  &   0.78  & 3.1 &   0.37  &   0.43  &   87.7  &   4.1   &   0.64  &   1.8  & --  & --  &  1.4  &   310   & (iii)\\
ALMA8 & 1.1 & 0.98 &   0.48  & 4.0 &  --  &  --  &  --  & -- &  --  &  --  &   140   &   245  &    7.2   &   37    & (i)\\
sub1  & 2.3 & 2.4  &   0.27  & 0.58&  --  &  --  &  -- & -- &  --  &  --  & --  & --  &    1.0   &    140  & (iii)\\
sub2  & 2.6 & 2.0  &   0.35  & 1.1 &  --  &  --  &  -- & -- &  --  &  --  & --  & --  &   6.5   &   30    & (iii)\\
sub3  & 1.4 & 1.5  &   0.33  & 1.3 &  --  &  --  &  -- & -- &  --  &  --  & --  & --  &   2.9   &   62    & (iii)\\
\enddata
\tablenotetext{a}{$N_{\mathrm{H_2, peak}}$ corresponds to the total gas column density estimated from the peak flux ($F_\mathrm{1.3 mm, peak}$) measured at the continuum peak.}
\tablenotetext{b}{Classifications in Section~\ref{sec:evolution}; (i)protostellar cores, (ii)protostellar core candidates, and (iii)prestellar core candidates.}
\end{deluxetable*}
\end{rotatetable*}

\subsection{ Line detection and spatial distribution of deuterated molecules} \label{sec:chemistry}
\begin{figure}
    \plotone{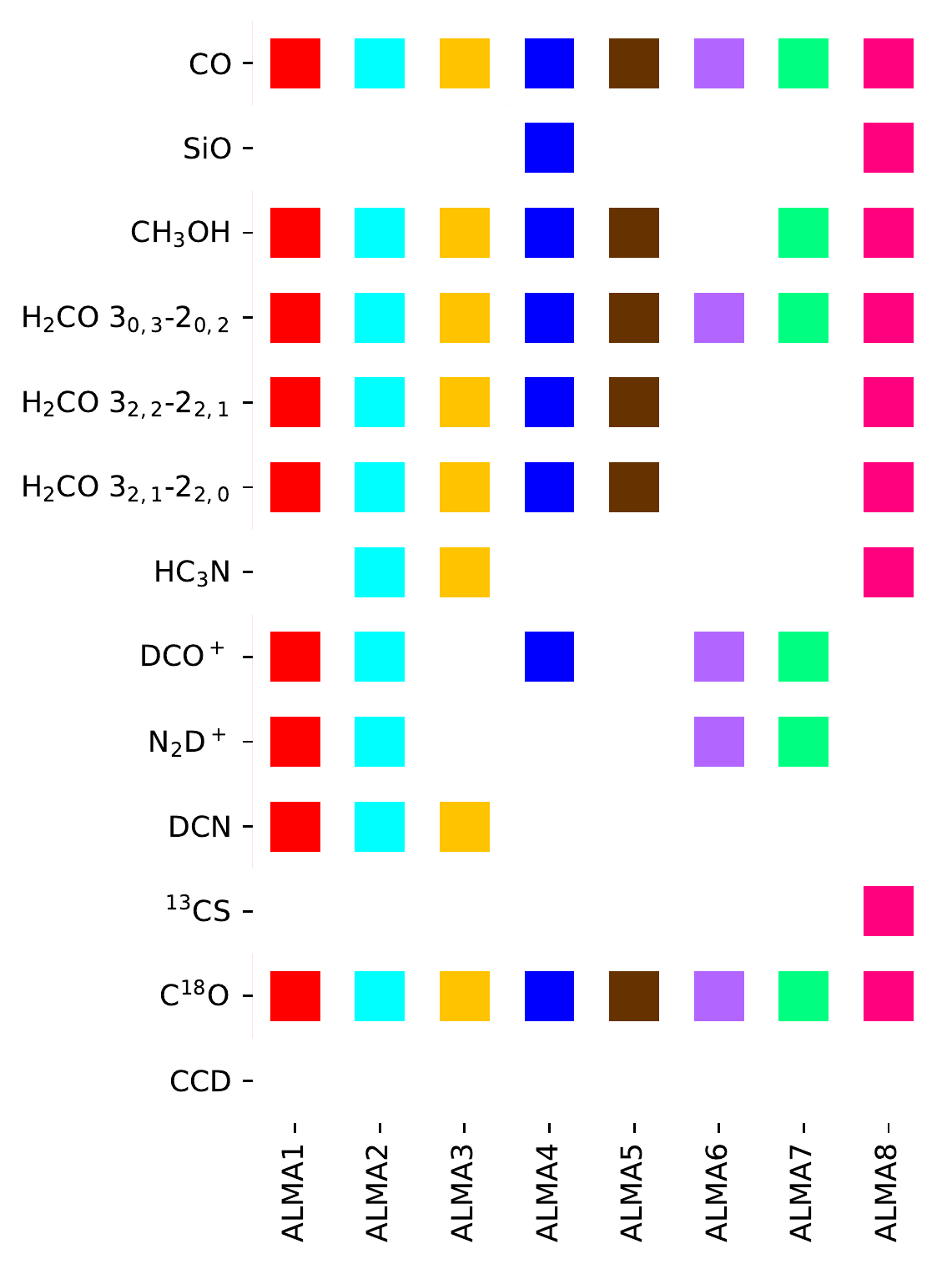}
    \caption{The summary of molecular detection in each sources. The detection limit was set as 3$\sigma$ at the continuum peak position. The order of molecule is same as that in Figure~\ref{fig:outflow_tracers} and ~\ref{fig:densegas_mom0}.
    }
    \label{fig:detection}
\end{figure}
We summarized the detection of molecular line emission in ALMA1--8 in Figure~\ref{fig:detection}. 
We defined the detection if the emission peak at the continuum peak position is brighter than 3$\sigma$, where  $\sigma$ is the rms measured in line-free channels (Table~\ref{tab:spw}). 
Spectra of deuterated molecules in addition to $^{13}$CS, C$^{18}$O, SiO, and CO are shown in Figure~\ref{fig:ave_profile}--\ref{fig:CO_profile} in Appendix. 
They are averaged within the core areas (ALMA1--ALMA8) identified by the dendrogram (Section \ref{sec:identification}).
\begin{figure*}
    \plotone{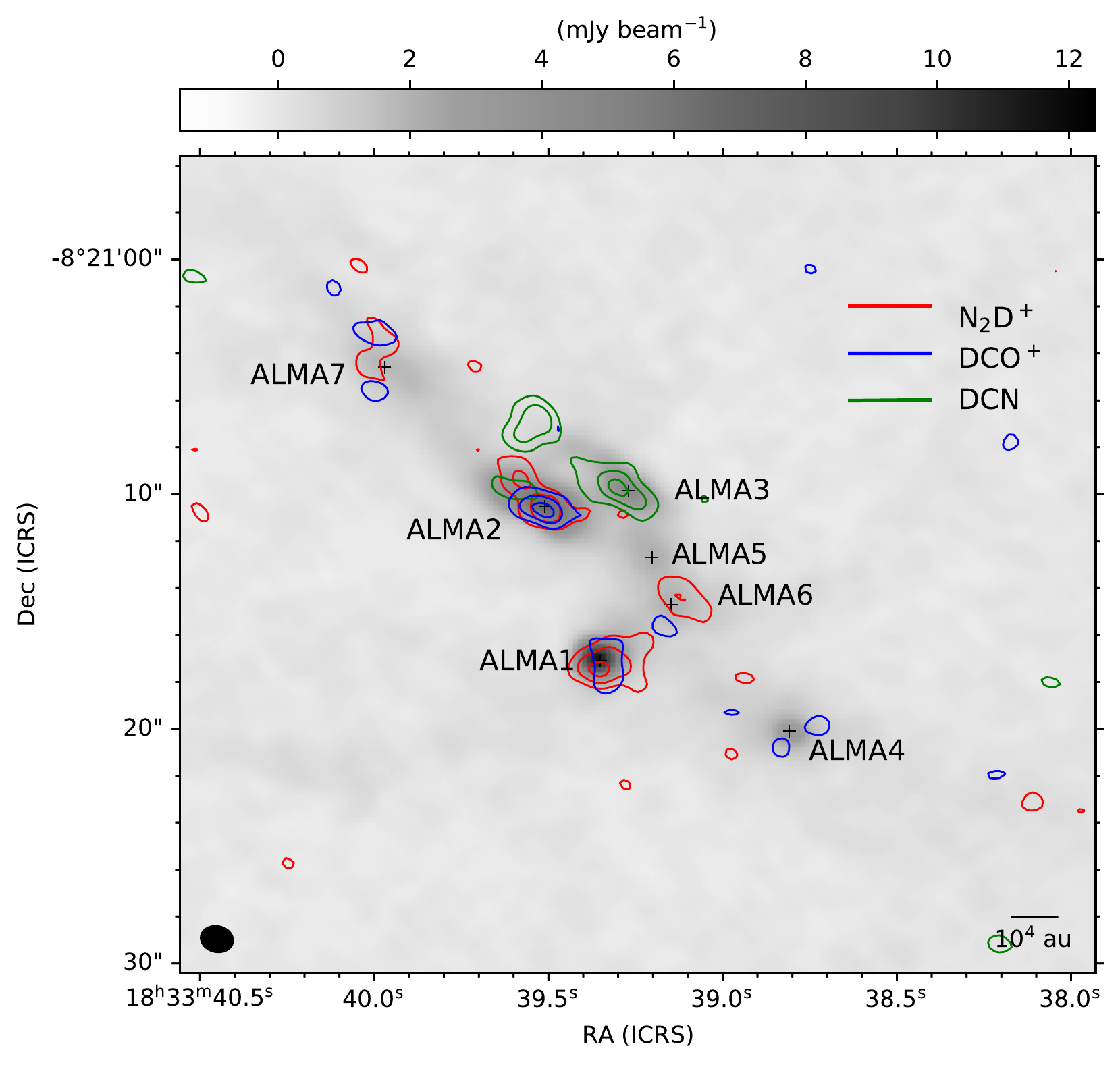}
    \caption{Integrated intensity map (moment 0) of N$_2$D$^+$ ($J$=3--2), DCO$^+$ ($J$=3--2) and DCN ($J$=3--2) overlaid with continuum emission.
    The red, blue, green contours correspond to N$_2$D$^+$, DCO$^+$, and DCN, respectively. The contour levels are 3, 5, 7, 10$\sigma_\mathrm{int}$, where $\sigma_\mathrm{int}$ is the rms of the integrated intensity map (1$\sigma_\mathrm{int}$ = 14, 10, and 8.9 mJy\,beam$^{-1}$, respectively).
    The gray scale shows the continuum emission.
    The black crosses correspond to the continuum peak of each core.
    The spatial scale and the beam size are shown at the bottom.}
    \label{fig:deuterated}
\end{figure*}

Figure \ref{fig:deuterated} shows the distribution of three dense gas tracers (N$_2$D$^+$, DCO$^+$, and DCN) overlaid with the dust continuum emission. 
Their spatial distribution is slightly different with each other, implying that these deuterated molecules seem to trace, at some degree, different environments.
The brightest N$_2$D$^+$ emission coincides with 
the continuum peak of ALMA1, and DCN emission  coincides with the continuum peak of ALMA3.

At an early stage of evolution prior to protostellar formation, 
molecules can be highly deuterated
in cold, dense regions because of freeze out of CO molecules onto dust grains under low temperatures \citep[$<$20 K; e.g.,][]{Caselli02}. 
In particular, the N$_2$D$^+$ molecule is destroyed by CO  \citep{jorgensen04,Salinas2017}, though DCO$^+$ and DCN molecules are not strongly affected by CO sublimation \citep{Turner01}.
In cold dense regions, DCN is likely to be depleted onto dust grains and sublimated at a temperature $\sim$50 K \citep[][]{Garrod17}. 
To detect DCN with high signal-to-noise, a warm region is necessary \citep[][]{Feng19}.
In fact, recently, Sakai et al. (2021, in prep.) study in detail the deuterated chemistry in IRDC G14.49, one of the ASHES sources from the pilot survey. 
They report that N$_2$D$^+$ emission traces quiescent regions, while DCO$^+$ and DCN emission trace active star-forming regions inside the IRDC. The difference in the spatial distribution of these three deuterated molecules may come from the different formation and destruction processes which are closely related to the environment. 

\subsection{Virial analysis}\label{sec:virial}
To investigate the stability of cores, we estimated virial masses following \citet{Liu20}.
The total virial mass accounting for both the magnetic field and the kinetic motions is given by
\begin{equation}
    M_\mathrm{k+B} = \sqrt{M^2_\mathrm{B}+\left(\frac{M_\mathrm{k}}{2}\right)^2} + \frac{M_\mathrm{k}}{2}.
\end{equation}
We omitted the contribution of external pressure. 
The kinetic virial mass and magnetic virial mass can be estimated from 
\begin{equation}
    M_\mathrm{k} = \frac{3(5-2a)}{3-a} \frac{R\sigma^2_\mathrm{tot}}{G}
\end{equation}
and 
\begin{equation}
    M_\mathrm{B} = \frac{\pi R^2 B_\mathrm{mag}}{\sqrt{\frac{3(3-a)}{2(5-2a)}\mu_0\pi G}},
\end{equation}
respectively, where $a$ is the index of the density profile ($\rho \propto r^{-a}$), $R$ is the radius of the core, 
$G$ is the gravitational constant, 
$B_\mathrm{mag}$ is the magnetic field strength, and $\mu_0$ is the permeability of vacuum. 
$\sigma_\mathrm{tot}$=$\sqrt{\sigma^2_\mathrm{th}+\sigma^2_\mathrm{nt}}$ is the total gas velocity dispersion. The thermal velocity dispersion and the non-thermal velocity dispersion are given by 
\begin{equation}
    \sigma^2_\mathrm{th} = \frac{kT}{\mu_\mathrm{p} m_\mathrm{H}}
\end{equation}
and
\begin{equation}
    \sigma^2_\mathrm{nt} = \sigma^2_\mathrm{DCO^+} - \frac{kT}{m_\mathrm{DCO^+}},
\end{equation}
respectively, where $\mu_\mathrm{p}$=2.33 is the conventional mean molecular weight per free particle considering H, He, and a negligible admixture of metals \citep[][]{Kauffmann08}. We assumed that the non-thermal component is independent of the molecular tracer and that $\sigma_\mathrm{DCO^+}$ is the observed velocity dispersion estimated by a Gaussian fitting to the DCO$^+$ profiles averaged within identified core areas ($m_\mathrm{DCO^+}$ is the mass of the DCO$^+$ molecule).
The ratio of the virial mass to the total gas mass derived using the continuum emission, known as the virial parameter, is defined as $\alpha_\mathrm{k+B}$ (= $M_\mathrm{k+B}/M_\mathrm{core}$). 

\begin{figure*}
    \plotone{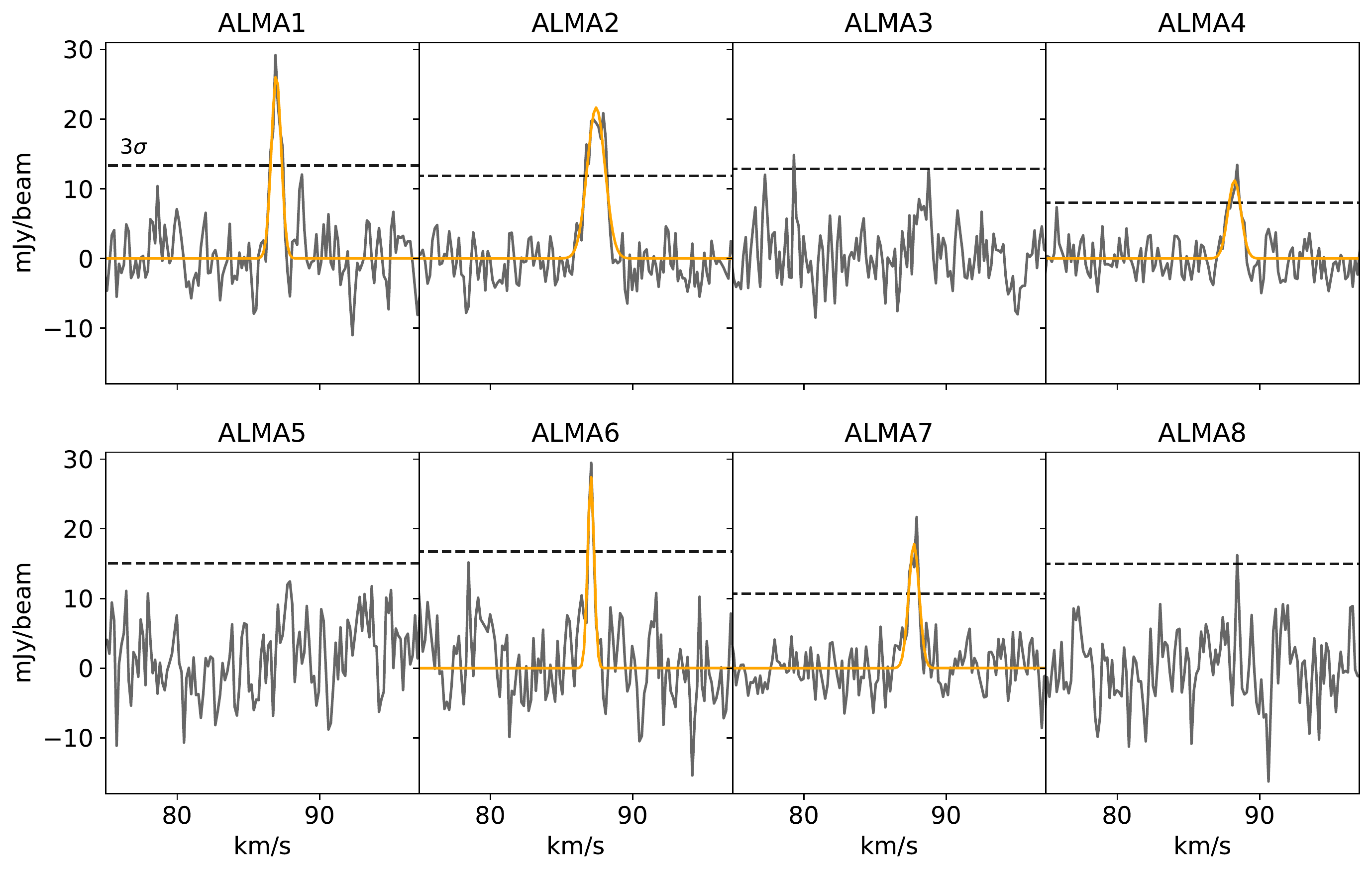}
    \caption{Spectra of DCO$^+$ ($J$=3--2) toward ALMA1--ALMA8 (grey) averaged within core areas identified by the dendrogram algorithm.  
    The horizontal dashed lines represent 3$\sigma_\mathrm{ave}$, where $\sigma_\mathrm{ave}$ is estimated in the averaged spectrum produced for each core (number of pixels averaged is different, so the  $\sigma_\mathrm{ave}$ value is also different per core). The orange lines show the results of the single Gaussian fitting. The parameters derived from the fitting results ($\sigma_\mathrm{DCO^+}$, $v_\mathrm{lsr}$) are summarized in Table~\ref{tab:phypara}.}
    \label{fig:DCO_prof}
\end{figure*}

Figure~\ref{fig:DCO_prof} shows the line spectra and the fitting results. The fitting succeeded for ALMA1, ALMA2, ALMA4, ALMA6, and ALAM7, where the amplitude of fitting result is larger than 3$\sigma$. 
Although we also obtained fitting results for N$_2$D$^+$ toward five cores, some N$_2$D$^+$ profiles were complex, likely due to the unresolved hyperfine structure of N$_2$D$^+$. We finally adopted the fitting results of the DCO$^+$ emission for the virial analysis. 
Table \ref{tab:phypara} lists $\sigma_\mathrm{DCO^+}$, $\sigma_\mathrm{tot}$, and the central velocity ($v_\mathrm{core}$) obtained from the Gaussian fitting for each core. 
We adopt the magnetic field strength $B_\mathrm{mag}$ = 2.6 mG, which is the average magnetic field strength estimated in three cores in G023.477 by using the Davis-Chandrasekhar-Fermi method \citep[][]{Beuther18}.
They conducted ALMA observations with an angular resolution 1.01$\arcsec \times 0.83\arcsec$, comparable to our observations. 

As listed in Table \ref{tab:phypara}, with the assumption that the density profile of the cores is uniform ($a$=0), $\alpha_\mathrm{k+B}$ ranges from 0.47 to 1.8.
Thus, ALMA4, ALMA6 and ALMA7 would be gravitationally supported by magnetic field. However, the massive cores of ALMA1 and ALMA2 are still unstable even by taking into account the magnetic field.
If the radial density profiles is not uniform (i.e., $a>0$), the virial parameter becomes smaller, indicating most cores are sub-virialized. 
For example, in the case of $a$=1.5, both the virial mass and virial parameter, with and without the contribution from the magnetic field are 0.87 and 0.80 times smaller, respectively.

\subsection{Tracers of warm gas}\label{sec:high-ex}
\begin{figure*}
\epsscale{1.2}
    \plotone{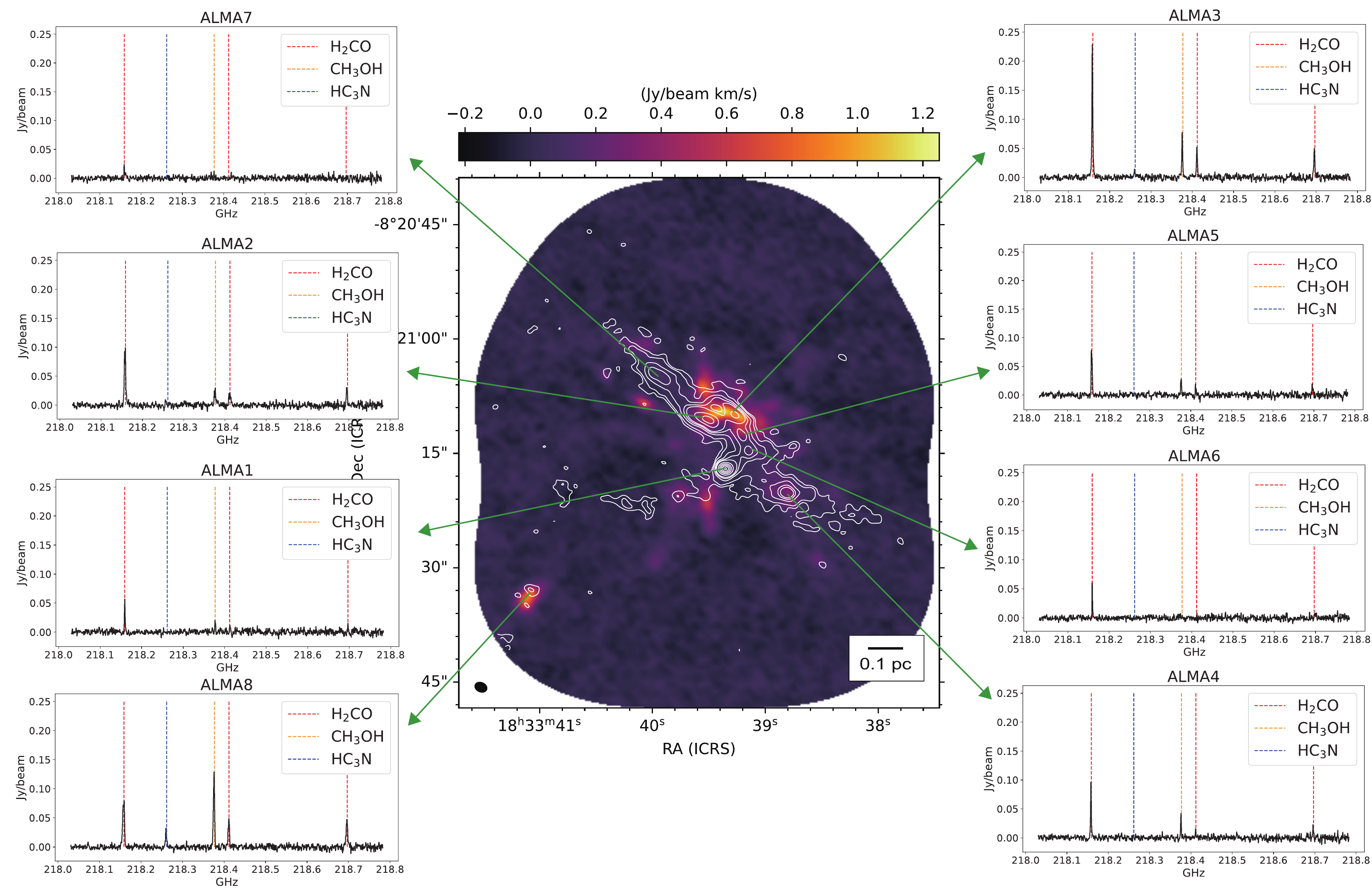}
    \caption{Integrated intensity map of H$_2$CO ($J =3_{0,3}$--$2_{0,2}$) overlaid with 1.3 mm continuum emission. The contour levels are consistent with Figure \ref{fig:continuum}. 
    The H$_2$CO beam size is plotted in the bottom left, and the spatial scale is in the bottom right.
    The panels around H$_2$CO image show line spectra including H$_2$CO ($J =3_{0,3}$--$2_{0,2}$), H$_2$CO ($J =3_{2,1}$--$2_{2,0}$), H$_2$CO ($J =3_{2,2}$--$2_{2,1}$), CH$_3$OH ($J =4_2$--$3_1$), and HC$_3$N ($J$=24--23).
    The orange lines correspond to the rest-frequency of H$_2$CO, the red ones corresponds to that of CH$_3$OH, and the blue one represents that of HC$_3$N.}
    \label{fig:H2CO_line}
\end{figure*}

In our observation,  
three H$_2$CO transition lines $J_\mathrm{K_a,K_c} = 3_{0,3}$--$2_{0,2} (E_\mathrm{u}/k = 20.96\,\mathrm{K})$, $J_\mathrm{K_a,K_c} = 3_{2,2}$--$2_{2,1} (E_\mathrm{u}/k = 68.09\,\mathrm{K})$ and $J_\mathrm{K_a,K_c} = 3_{2,1}$--$2_{2,0} (E_\mathrm{u}/k = 68.11\,\mathrm{K})$, one CH$_3$OH transition line $J_\mathrm{K} = 4_2$--$3_1 (E_\mathrm{u}/k = 45.46\,\mathrm{K})$, and HC$_3$N (v=0, $J$=24--23, $E_\mathrm{u}/k = 131\,\mathrm{K}$) are detected toward several cores. 
Figure~\ref{fig:H2CO_line} shows these spectral lines at the continuum peak of each core.  
The red dashed vertical lines correspond to the H$_2$CO transitions, the orange ones correspond to CH$_3$OH, and the blue ones represent HC$_3$N.
If the detection limit is set at $3\sigma$ (1$\sigma$=2.76 mJy beam$^{-1}$), all five lines are detected only from ALMA3 and ALMA8. All lines except HC$_3$N are detected from ALMA1, ALMA2, ALMA4, and ALMA5. From ALMA6 and ALMA7, only the H$_2$CO $(3_{0,3}$--$2_{0,2})$ line is detected.

H$_2$CO line emission has been used to measure the gas temperature \citep[e.g.,][]{Tang17,Lu17}. Using the rotational diagram technique, we estimated the H$_2$CO rotation temperature at the dust peak position by fitting a single Gaussian component to the three transitions, following \citet{Turner91}. 
With the assumption of LTE and optically thin conditions, the relationship among the column density ($N_\mathrm{total}$), the rotation temperature ($T_\mathrm{rot}$), and the brightness temperature ($T_\mathrm{B}$) is described as
\begin{equation}
    \mathrm{ln} L = 
    \mathrm{ln} \left( \frac{N_\mathrm{total}}{Q(T_\mathrm{rot})}\right)-
    \frac{E_\mathrm{u}}{k}\frac{1}{T_\mathrm{rot}},
\end{equation}
where 
\begin{equation}
    L = \frac{3k\int T_\mathrm{B}dv}{8\pi^3\nu S \mu^2 g_\mathrm{I}g_\mathrm{K}},
\end{equation}
and 
\begin{equation}
    \int T_\mathrm{B}dv = \left(2\sqrt{\frac{\ln 2}{\pi}}\right)^{-1}T_\mathrm{B,peak}\Delta v_\mathrm{H_2CO}.
    \label{equ:gaussianfit}
\end{equation}
Here, $E_\mathrm{u}, S, \mu, g_\mathrm{I}$, $g_\mathrm{K}$, and $\Delta v_\mathrm{H_2CO}$ are the upper state energy, the line strength, the relevant dipole moment, the reduced nuclear spin degeneracy, the $K$-level degeneracy, and the FWHM of the corresponding H$_2$CO line. Equation~(\ref{equ:gaussianfit}) represents the relation for the area of a Gaussian with a peak brightness temperature ($T_\mathrm{B,peak}$) and a FWHM.

The partition function $Q(T_\mathrm{rot})$ is approximated as 
\begin{equation}
    Q(T_\mathrm{rot})\sim \frac{1}{2}\left[\frac{\pi(kT_\mathrm{rot})^3}{h^3ABC}\right]^{1/2}~, 
\end{equation}
where $A=281.97037\,\mathrm{GHz}, B=388.354256\,\mathrm{GHz},$ and $C=340.057303\,\mathrm{GHz}$ are the rotational constants.
The peak brightness temperature $T_\mathrm{B,peak}$, in K, was calculated from the peak intensity $S_\mathrm{peak}$, in Jy beam$^{-1}$, as
\begin{equation}
    T_\mathrm{B,peak} =\frac{c^2}{2k\nu^2}S_\mathrm{peak}\Omega.
\end{equation}
The estimated H$_2$CO column density and rotation temperature are listed in Table \ref{tab:phypara}, and the rotational diagrams are shown in Appendix (Figure~\ref{fig:rot_diagram}).
At $T_\mathrm{rot}$\,=\,245 K, ALMA8 has the highest temperature among all cores. 
Relatively massive cores, ALMA1-4, have similar rotational temperatures, ranging between 43 and 62 K. 
ALMA5 has the lowest temperature, $T_\mathrm{rot}\sim$ 37\,K. 
For ALMA6 and ALMA7, we did not detect the two H$_2$CO transition lines ($3_{2,2}$--$2_{2,1}$ and $3_{2,1}$--$2_{2,0}$). Therefore, for these cores, we derived the upper limits of the rotational temperatures as 63\,K, assuming the 3\,$\sigma$ intensity strengths with the average line widths (1.75\,km\,s$^{-1}$) among other cores for these lines. 

To derive the rotational temperatures, we assumed that all three H$_2$CO lines are optically thin.  To check the validity of this assumption, we derive the optical depths of the lines using the RADEX\footnote{http://var.sron.nl/radex/radex.php} non-local thermodynamical equilibrium model \citep{vandertak07}. 
Using the derived rotation temperature and column density of H$_2$CO, the number density of the H$_2$ gas, and the velocity dispersion of H$_2$CO ($\sim$3-5\,km\,s$^{-1}$),
the optical depths are estimated as a few $\times 10^{-3}$, except for ALMA8. Thus, our assumption of the optically thin condition is appropriate for all cores, except one. In the case of ALMA8, the H$_2$CO emission is likely optically thick, resulting in an overestimation of the derived temperature by using the rotational diagram technique.  

It is worth noting that the distribution of the H$_2$CO emission resembles that of the SiO emission, indicating that the H$_2$CO emission is affected by protostellar activity (such as outflows). \cite{Tang17} find that in regions associated with molecular outflows or shocks, the temperature derived from H$_2$CO is distinctly higher than temperatures derived from NH$_3$ or dust emission. They also find that the turbulence traced by H$_2$CO is higher than that traced by other typical tracers of quiescent gas, such as NH$_3$. Here in G023.477, we find that line widths of H$_2$CO are also larger than those of the dense gas tracers such as DCO$^+$ and N$_2$D$^+$, typically by a factor 4. Therefore, it is highly likely that H$_2$CO does not represent well the core  kinematics nor their temperature, consequently the rotational temperature is not assumed for the determination of core physical parameters. More details on the H$_2$CO emission of the whole ASHES sample will be presented in Izumi et al. (2021, in prep.).

\subsection{C$^{18}$O depletion}

Since low temperature and high density conditions allow CO to freeze out onto dust grains, low abundances of CO and its isotopologues can be used as indicators of cold and dense  regions. 
In this subsection, to investigate such cold regions without active star formation,
we estimate the integrated C$^{18}$O depletion factor, $f_\mathrm{D}$, which is defined as the ratio between the expected (i.e., canonical) abundance of C$^{18}$O relative to H$_2$, $X^\mathrm{E}_\mathrm{C^{18}O}$, and the abundance estimated from observed value, $X_\mathrm{C^{18}O}$ as
\begin{equation}
    f_\mathrm{D} = \frac{X^\mathrm{E}_\mathrm{C^{18}O}}{X_\mathrm{C^{18}O}},
\end{equation}
where $X_\mathrm{C^{18}O}$ is the ratio of the observed C$^{18}$O column density ($N_\mathrm{C^{18}O}$) to the observed H$_2$ column density ($N_\mathrm{H_2, peak}$) derived from continuum emission.

Assuming that C$^{18}$O ($J$=2--1) is optically thin and under LTE condition, we derived the column density of C$^{18}$O by adopting the dust temperature of 13.8 K as the excitation temperature  ($T_\mathrm{ex}$).
 We fitted the C$^{18}$O emission at the continuum peak of each core with a single Gaussian.
With the assumption mentioned above, the column density is derived by using the following equation \citep[][]{Mangum15,Sanhueza12}:
\begin{eqnarray}
    N &=& \frac{3h}{8\pi^3\mu^2J_u}\left(\frac{kT_\mathrm{ex}}{hB_\mathrm{C^{18}O}}+ \frac{1}{3}\right)\,\frac{\mathrm{exp}(E_\mathrm{u}/kT_\mathrm{ex})}{\mathrm{exp}(h\nu/kT_\mathrm{ex})-1}\nonumber\\
    && \times \frac{\int T_\mathrm{B}dv}{J(T_\mathrm{ex})-J(T_\mathrm{bg})},
\label{equ:columndensity}
\end{eqnarray}
where $B_\mathrm{C^{18}O}$ is the rotational constant of C$^{18}$O, 54.891421 GHz, $J_u$ is the rotational quantum number of the upper state, and $J(T)$ is defined by
\begin{equation}
    J(T) = \frac{h\nu}{k} \frac{1}{\mathrm{exp}(h\nu/kT)-1}.
\end{equation}. 

The expected CO abundance at the galactocentric distance $R_\mathrm{GC}$ is calculated using the relationship \citep[][]{Fontani06} as
\begin{equation}
    X^E_\mathrm{CO} = 9.5 \times10^{-5}\,e^{1.105-0.13R_\mathrm{GC}\mathrm{[kpc]}}.
\end{equation}
To calculate the expected C$^{18}$O abundance, we take into account the dependence of the oxygen isotope ratio $[^{16}\mathrm{O}]/[^{18}\mathrm{O}]$ on $R_\mathrm{GC}$ according to \citep[][]{Wilson94}
\begin{equation}
    \frac{[^{16}\mathrm{O}]}{[^{18}\mathrm{O}]} = 58.8 \times R_\mathrm{GC}\mathrm{[kpc]}+37.1.
\end{equation}
Finally, the expected C$^{18}$O abundance is obtained as
\begin{eqnarray}
    X^E_\mathrm{C^{18}O} &=& \frac{X^E_\mathrm{CO}}{[^{16}\mathrm{O}]/[^{18}\mathrm{O}]}\nonumber\\
    &=& \frac{9.5 \times 10^{-5}\,e^{1.105-0.13R_\mathrm{GC}\mathrm{[kpc]}}}{58.8 \times R_\mathrm{GC}\mathrm{[kpc]}+37.1}.
\end{eqnarray}
Table~\ref{tab:phypara} lists the calculated column density of C$^{18}$O ($N_\mathrm{C^{18}O}$) and the depletion factor ($f_\mathrm{C^{18}O}$) for each core. 
While most cores have a depletion factor around 60, as expected for IRDCs, ALMA1 and ALMA7 have significantly higher values ($>$300), suggesting these cores are likely the coldest and have not been much affected by star formation activity (such as heating and outflows). 
Such difference comes from the largely different C$^{18}$O abundances ($X_\mathrm{C^{18}O}$) among cores. Our analysis shows the C$^{18}$O abundance vary with a factor of $\sim$10 in the same cloud.
The estimated depletion factors ($f_\mathrm{C^{18}O}$) are higher on average than evolved high-mass star forming region using single-dish observations \citep[e.g.,  $<$15;][]{Feng20} but comparable to that estimated in a core located in another IRDC G028.37+00.07-C1 using interferometric observations \citep[$>$616;][]{Kong18}.

The estimated core densities are as high as 10$^6$\,cm$^{-3}$, and thus the C$^{18}$O lines could have optical depths of $\tau\gtrsim$1.
Considering the effect of the optical depth, the column density (\ref{equ:columndensity}) is multiplied by a factor of $\tau/ (1-e^{-\tau})$.  If the optical depth is as high as $\tau\sim$5, the C$^{18}$O column densities become 5 times larger, resulting in the 5 times smaller depletion factor.
\section{Outflows} \label{sec:outflow}

\begin{figure*}
    \plotone{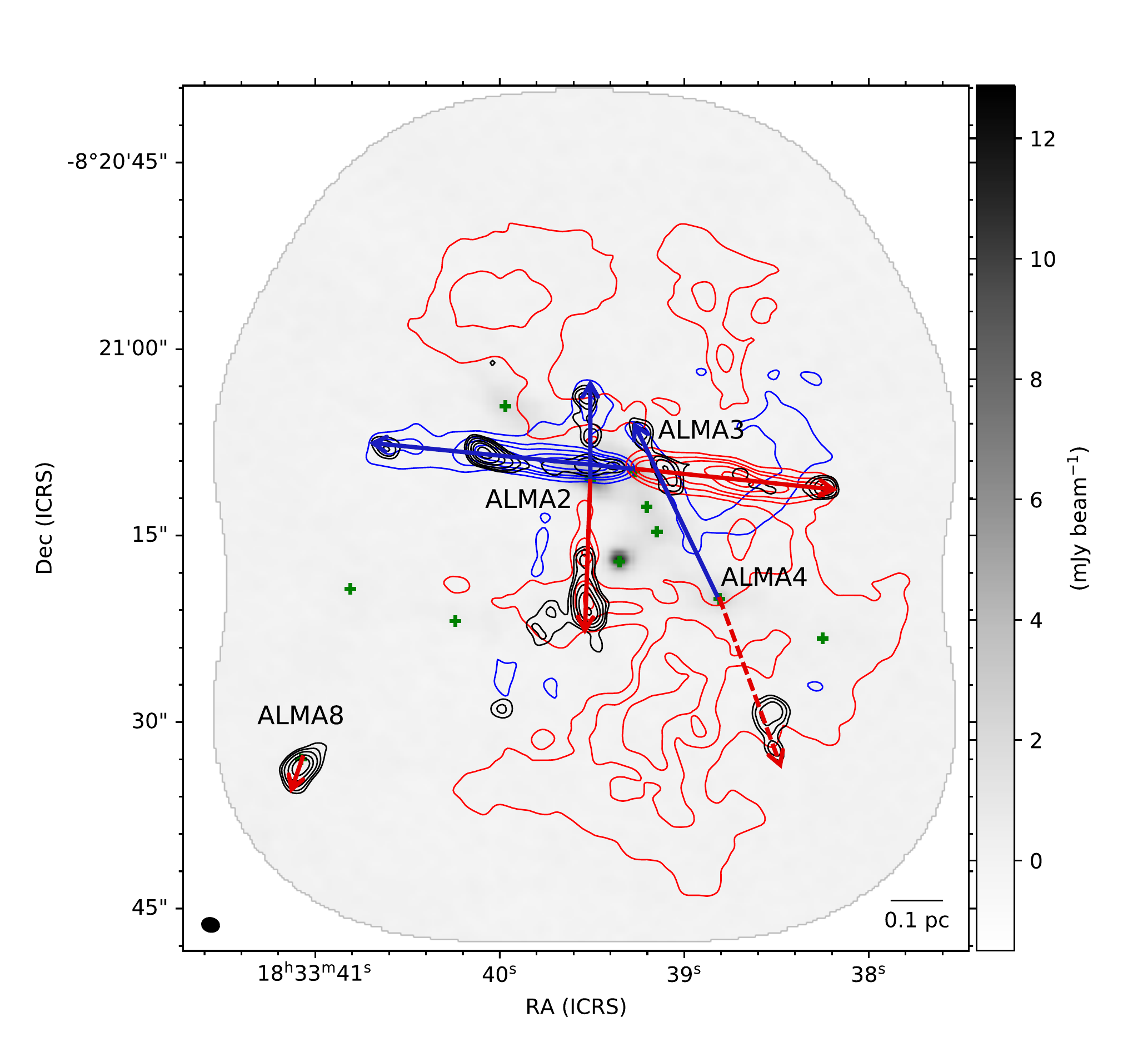}
    \caption{
    The grayscale image is the continuum emission same as Figure \ref{fig:continuum}.
    The blue and red contours show the integrated intensity of blue-shifted and red-shifted CO emission, respectively. The blue-shifted component is integrated from 20.4 km\,s$^{-1}$ to 77.4 km\,s$^{-1}$, and red-shifted component is integrated from 97.4 km\,s$^{-1}$ to 180.4 km\,s$^{-1}$.
    The black  contours represent SiO integrated from 46.7 km\,s$^{-1}$ to 126 km\,s$^{-1}$.
    Contour levels are set 4, 7, 10, 15, 20, and  30$\sigma_\mathrm{int}$ (1$\sigma_\mathrm{int}$ = 0.32, and 0.043 Jy beam$^{-1}$ for CO and SiO, respectively).
    The green ``+" symbols are the peak positions of the continuum emission.
    The black ellipse in the bottom left corner shows the synthesized beam size.
    The spatial scale is shown in the bottom right.}
 \label{fig:outflow} 
\end{figure*}
\subsection{Outflow identification}

CO ($J$=2--1) and SiO ($J$=5--4) are useful outflow and shock tracers. 
As mentioned in Section~\ref{sec:line_emission}, at least two collimated structures can be seen in both CO and SiO integrated intensity maps.
To search for high-velocity components which are likely to originate from outflows,
we examine the CO cube and the integrated intensity maps for blue- and red-shifted components separately. 
Figure~\ref{fig:outflow} shows the blue- and red-shifted components of CO and SiO emission overlaid on the continuum image. 
The CO and SiO line emission unveiled outflows ejected from ALMA2, ALMA3, ALMA4, and ALMA8, though the red-shifted outflow from ALMA4 cannot be separated from the ambient gas.
No outflow is detected from ALMA1, which has the highest peak intensity in this region.
Since ALMA8 is located at the edge of the field-of-view and CO  intensity is low, we can see only SiO emission in the integrated intensity map. 

\subsection{Outflow parameters} \label{sec:outflow-para}
We define the outflow components by using the CO ($J$=2--1) data cube and the integrated intensity map following \citet{Li19a,Li20}. 
Based on the region where CO is brighter than the 4$\sigma$ noise level in p-p-v space, we determined the intrinsic maximum outflow velocity ($\Delta v_\mathrm{max}$ = $|v_\mathrm{LSR} - v_\mathrm{sys}|$), where $v_\mathrm{sys}=86.5$\,km\,s$^{-1}$ as mentioned in Section~\ref{sec:virial}.
The maximum projected distance ($\lambda_\mathrm{max}$) is defined from the CO emission above 4$\sigma_\mathrm{int}$ in the integrated intensity map, though we used SiO emission for an outflow associated from ALMA8.
This $\sigma_\mathrm{int}$ = 0.32 Jy beam$^{-1}$ is the rms noise level measure in the integrated intensity map.
The maximum outflow velocity ranges between 12 and 94 km\,s$^{-1}$, and the projected outflow length for each lobe varies from 0.17 to 0.50 pc.
ALMA3 has the longest ($\lambda_\mathrm{max,b}+\lambda_\mathrm{max,r}$ =  0.87 pc) and the fastest ($\Delta v_\mathrm{max, b}$ = 66\,km\,s$^{-1}$ and $\Delta v_\mathrm{max, r}$ = 94\,km\,s$^{-1}$) outflow. 
The subscripts ``b'' and ``r'' indicate ``blue-'' and ``red-'' shifted components, respectively.
We also independently measured the outflow position angles for both the blue- and red-shifted lobes by connecting the continuum peak with the peak of the integrated intensity maps of CO emission.
The measured angles range from -94\arcdeg\ to +180\arcdeg\ counterclockwise from the celestial North. 
All values are listed in Table \ref{tab:outflow}, and the channel map is shown in Appendix.

To estimate the dynamical timescale, we use the projected distance ($\lambda_\mathrm{max}$) and the maximum velocity without considering the inclination of the outflow axis with respect to the line of sight as 
\begin{equation}
    t_\mathrm{dyn} = \frac{\lambda_\mathrm{max}}{\Delta v_\mathrm{max}}.
    \label{equ:tdyn}
\end{equation}
Assuming LTE conditions and that the CO emission in the outflowing gas is optically thin, the CO column density ($N_\mathrm{CO}$) is derived from Equation~(\ref{equ:columndensity}).
The outflow mass ($M_\mathrm{out}$), momentum ($P_\mathrm{out}$), and energy ($E_\mathrm{out}$) are estimated as \citep{Bally83,Carbit92,Mangum15}: 
\begin{eqnarray}
    M_\mathrm{out} &=& d^2 \bar{m}_\mathrm{H_2}X_\mathrm{CO}^{-1}\,\int_\Omega N_\mathrm{CO} d\Omega, \label{equ:outflowmass}\\
    P_\mathrm{out} &=& M_\mathrm{out} \Delta v,\label{equ:outflowP}\\
    E_\mathrm{out} &=& \frac{1}{2}M_\mathrm{out} (\Delta v)^2\label{equ:outflowE}.
\end{eqnarray}
Here, 
$\Omega$ is the total solid angle that the flow subtends, $d$ is the source distance, and $\Delta v$ is the outflow velocity with respect to the systemic velocity ($\Delta v = |v_\mathrm{LSR}-v_\mathrm{sys}|$). 
In this work, we assumed that the excitation temperature of the outflow gas is 30 K, and adopt a CO-to-H$_2$ abundance ($X_\mathrm{CO}$) of 10$^{-4}$ \citep{Blake87}.
If we change the excitation temperature from 20 K to 60 K, the effect on the estimated column density is less than 50$\%$. 
Using the dynamical timescale derived from (\ref{equ:tdyn}), outflow mass (\ref{equ:outflowmass}), momentum (\ref{equ:outflowP}), and energy (\ref{equ:outflowE}), we compute the outflow rate
($\dot M_\mathrm{out}$), outflow luminosity ($L_\mathrm{out}$), and mechanical force ($F_\mathrm{out}$) as:
\begin{eqnarray}
    \dot M_\mathrm{out} &=& \frac{M_\mathrm{out}}{t_\mathrm{dyn}},\\
    L_\mathrm{out} &=& \frac{E_\mathrm{out}}{t_\mathrm{dyn}},\\
    F_\mathrm{out} &=& \frac{P_\mathrm{out}}{t_\mathrm{dyn}}.
\end{eqnarray}
The estimated outflow dynamical timescales range from $4.0 \times 10^3$ to $1.4 \times 10^4\,\mathrm{yr}$, and outflow masses range from 0.032 to 1.3\,$M_\odot$. 
The ejection rates are calculated between $2.2 \times 10^{-6}$ and 2.6 $\times 10^{-4}\,\dot{M}_\odot\,\mathrm{yr}^{-1}$.
All outflow parameters are summarized in Table \ref{tab:outflow}. 

\citet{Li20} reported the detection of 43 outflows in nine IRDCs from the ASHES pilot survey \citep{Sanhueza19}.  
As shown in Figure 3 of \citet{Li20}, the average maximum velocity was around 20 \,km\,s$^{-1}$, and the average maximum projected distance was around 0.17 pc. 
While the outflow parameters of ALMA2, ALMA4, and ALMA8  are similar to these values, the outflow of ALMA3 has higher values in both properties. ALMA3 has the most extreme properties so far discovered in the ASHES sample, being also the most massive and having the largest outflow mass rate. 

\begin{deluxetable*}{llcccccc}
\label{tab:outflow}
\tabletypesize{\footnotesize}
\tablecaption{Parameters of  Identified Outflows}
\tablewidth{0pt}
   \tablehead{ & &\multicolumn{2}{c}{ALMA2}& \multicolumn{2}{c}{ALMA3}&\colhead{ALMA4} & \colhead{ALMA8}  \\ & unit & blue& red & blue & red & blue & red}
   \startdata
   $\lambda_\mathrm{max}$ & pc & 0.18 & 0.28 & 0.50 & 0.37 & 0.37 & 0.17 \\
   $\Delta v_\mathrm{max}$ &km \,s$^{-1}$ &24& 36 & 66& 94&28&12 \\
    PA &  deg & 4 & 180 & 82 & -94 & 24 & 156 \\
   $t_\mathrm{dyn}$ & 10$^4$ yr & 0.77&0.80 & 0.78 &0.40& 1.4 & 1.5 \\
   $M_\mathrm{out}$ & $M_\odot$ & 0.25 & 0.27 & 0.61 & 0.70 & 0.50  & 0.032\\
   $P_\mathrm{out}$ & $M_\odot$ \,km \,s$^{-1}$ & 3.4 & 6.4& 11 & 19 & 7.1 & 1.0 \\
   $E_\mathrm{out}$ & 10$^{45}$ erg & 0.55 & 2.4 & 3.5 & 7.1 & 1.6 & 0.50 \\
   $\dot M_\mathrm{out}$ & 10$^{-5} M_\odot$ yr$^{-1}$ & 3.3 & 3.4 & 8.8 & 17 & 3.7 & 0.22 \\
   $F_\mathrm{out}$ &10$^{-4} M_\odot$\,km\,s$^{-1}$ yr$^{-1}$ & 4.4 & 8.0 & 22 & 46 & 5.2 & 0.71 \\
   $L_\mathrm{out}$ & 10$^{33}$erg s$^{-1}$ & 2.4 & 9.9 & 26 & 58 & 4.4  & 1.2
   \enddata
   \tablenotetext{}{The parameters are not corrected for the inclination angle of outflows ($\theta = 0$).}
\end{deluxetable*}

\subsection{PV diagrams}
\begin{figure*}
  \epsscale{1.0}
    \plotone{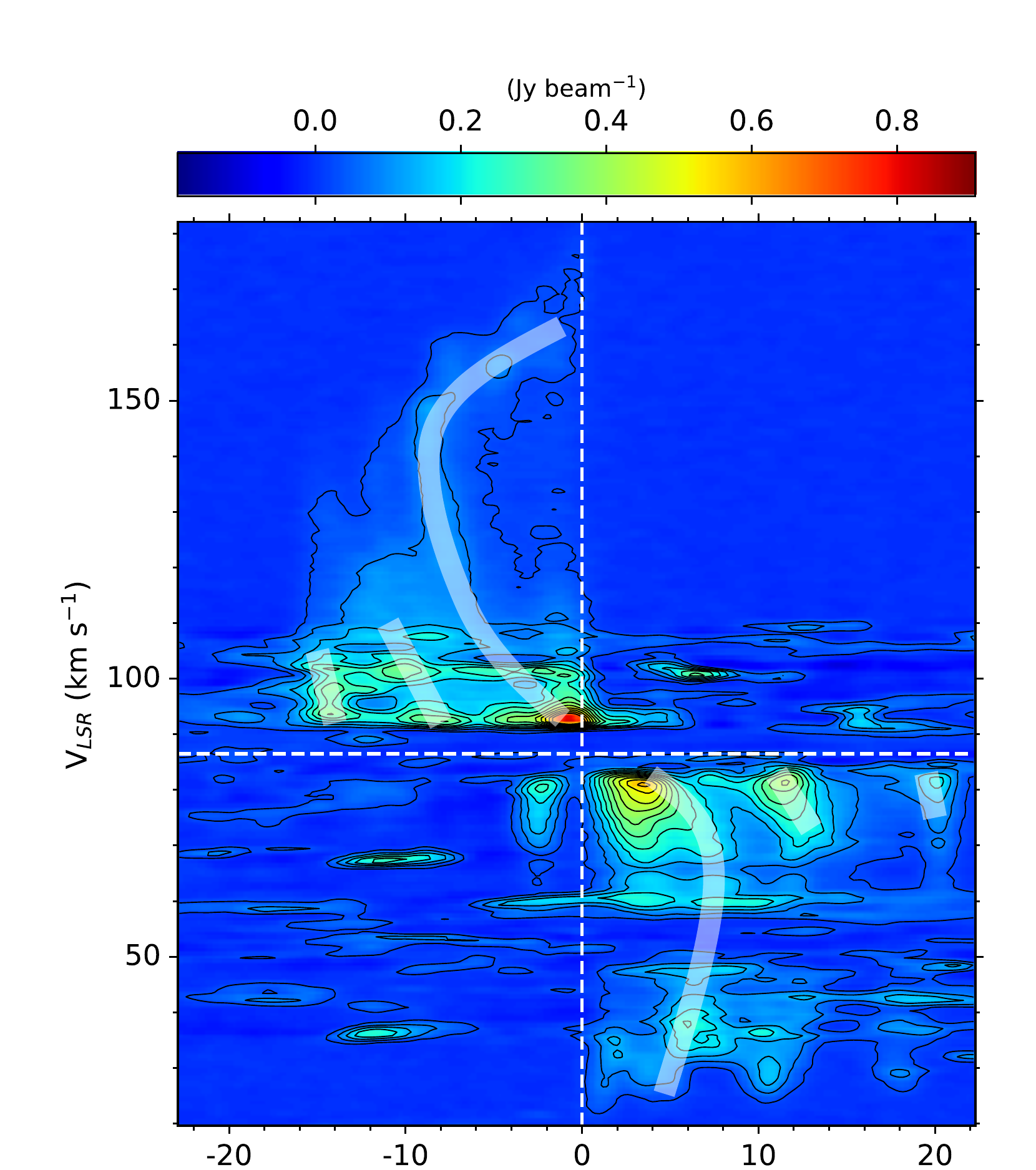}
    \caption{Position-Velocity (PV) diagram of CO emission for ALMA3. The cut of the PV diagrams is along the CO outflow (P.A. = $\sim$81\arcdeg) with width=3\,pix (1\,pix=0.2$''$). The contour levels are 3$\sigma$, 10$\sigma$ to 230$\sigma$ by 20$\sigma$ steps (1$\sigma$=2.64 mJy beam$^{-1}$).
    The white curved lines shows the S-shape structure, and white lines show the knotting structure called Hubble wedges.
    The vertical white dashed line is the position offset=0, and the horizontal one corresponds to the systemic velocity of this region. }
 \label{fig:alma3pv} 
\end{figure*}

\begin{figure*}
    \epsscale{1.1}
    \plotone{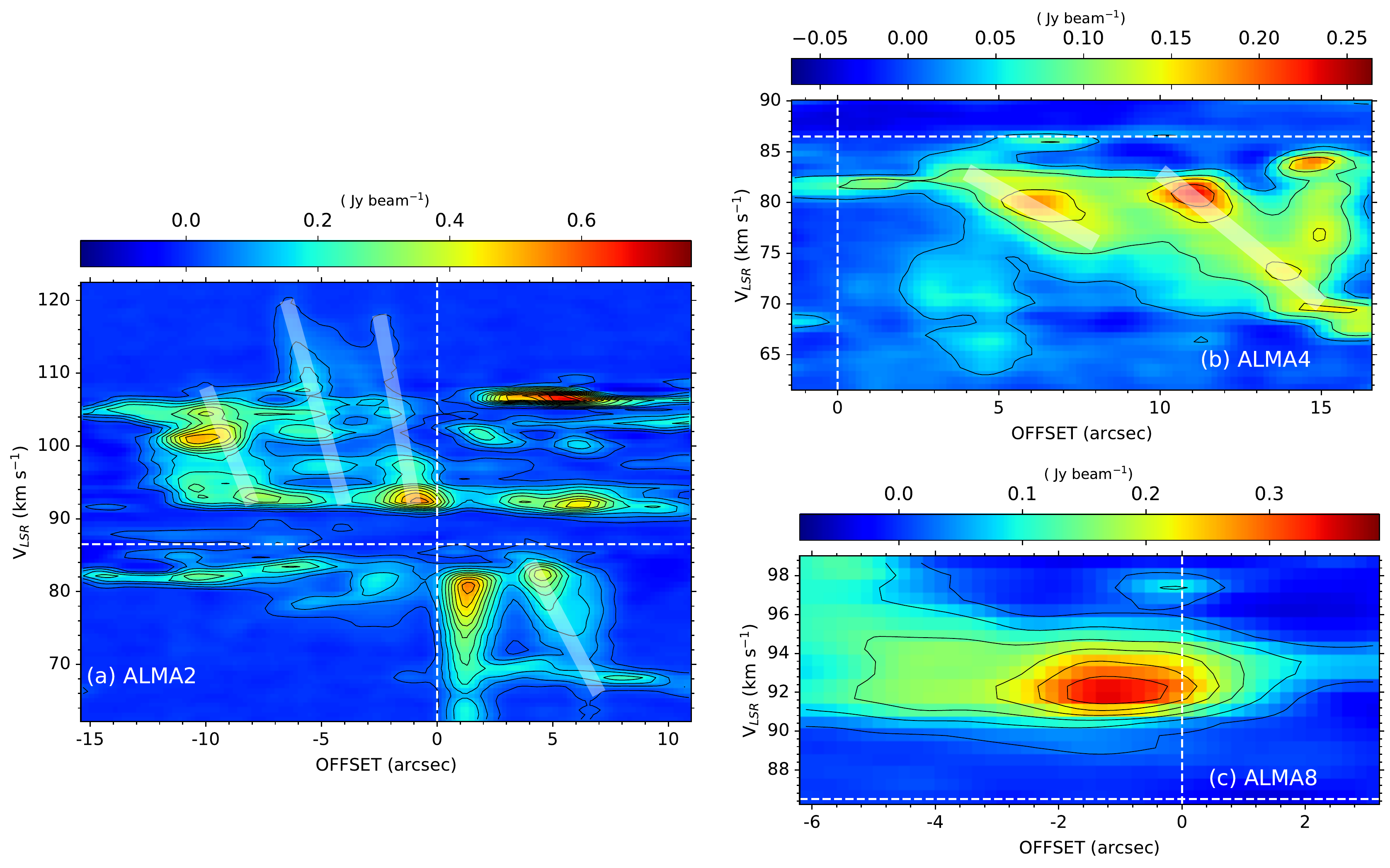}
    \caption{Position-velocity (PV) diagram of the CO emission associated with (a) ALMA2, (b) ALMA4, and (c) ALMA8. 
    The white lines in the panel (a) and (b) indicate Hubble Wedges, where the gas velocity increases with distance to the protostar.
    The vertical and horizontal white dotted lines show the position and velocity of the core, respectively. The contour levels are  3$\sigma$, 10$\sigma$ to 230$\sigma$ by 20$\sigma$ steps 
    (1$\sigma$ = 2.64 mJy\,beam$^{-1}$). }
    \label{fig:pv247}
\end{figure*}

The Position-Velocity (PV) diagram is useful to disentangle the ejection process of outflows. 
Figure \ref{fig:alma3pv} shows the PV diagram cut along the outflow ejected from ALMA3 (P.A. = $\sim$81\arcdeg).  
As denoted as white lines, we can confirm some knotting structures 
in the lower velocity region, $v_\mathrm{LSR}$= 50 -- 120\,km\,s$^{-1}$, in some of which the velocity increases with increasing distance from the core. 
Such structures are referred as Hubble wedges  \citep{ArceGoodman01}, indicating episodic mass ejection.   
In the higher velocity range area of the PV diagram, we can recognize a S-shape structure, which is indicated by thick white lines (Figure~\ref{fig:alma3pv}). 
The S-shape structure in the PV diagram consists of two components based on their slope in the PV diagram. 
One is a low-velocity component, whose velocity increases with increasing distance, and the other is a high-velocity component, whose velocity decrease with increasing distance.
\citet{Tafoya21-arxiv} firstly reported a similar peculiar S-shaped morphology in the PV diagram detected in IRDC G10.99-0.08 (part of the ASHES pilot survey). 
They explain such S-shape structures in the PV diagrams by two different gas components based on the jet-driven outflow scenario \citep[][]{Shang06}. The low-velocity component traces the gas entrained by a high-velocity jet and the high-velocity one is associated with the jet that moves with high velocity, but decelerates \citep[][]{Tafoya19, Tafoya21-arxiv}.
While the outflow from ALMA3 does not exhibit the exact S-shaped morphology seen in \citet{Tafoya21-arxiv}, because the episodic ejections, the outflow from ALMA3 is likely to be the second example showing S-shaped structure in the PV diagram in star-forming regions. Coincidentally, this second example is also found in a very young protostellar object embedded in a 70 $\mu$m dark IRDC, hinting that such shape in the PV diagram may preferentially  
appear at the very early stages of star formation, when the driving jet has a stronger interaction with the quiescent material of the ambient medium.

The PV diagrams of the other outflows associated with ALMA2, ALMA4, and ALMA8 are shown in Figure \ref{fig:pv247}.
All images indicate the gas velocity increases with distance to the protostar, which is called the Hubble Law. 
In particular, 
the PV diagram of ALMA2 (Figure~\ref{fig:pv247} (a)) show multiple Hubble Law wedges, 
which again indicates 
episodic accretion history \citep[][]{ArceGoodman01}. 
These features have been also observed in other IRDCs and in other active high-mass star-forming regions  \citep[e.g.,][]{Li20,Nony20}. 
The flaring of some high-mass protostars has been also observed in near-infrared \citep[][]{Caratti17}.
All these observations support the picture that an important fraction of protostars in high-mass star-forming regions undergo episodic accretion. 

\section{Discussion} \label{sec:discussion}
\subsection{Position angle of outflows} \label{sec:position}
\begin{figure}
    \centering
    \plotone{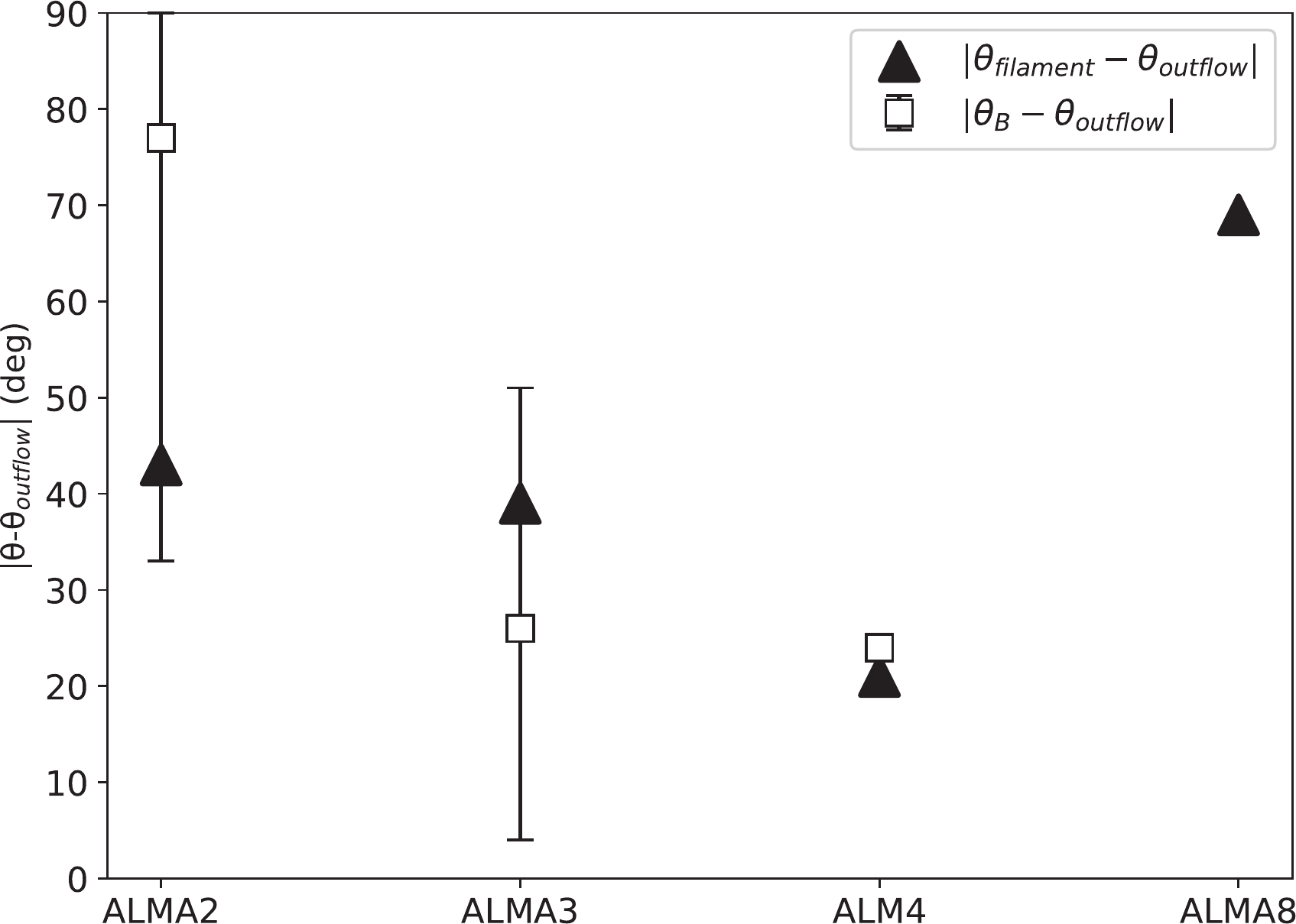}
    \caption{
    The projected separations of the outflow position angles ($\theta_\mathrm{outflow}$) with respect to magnetic field orientation ($\theta_\mathrm{B}$) and the filament ($\theta_\mathrm{filament}$) as open squares with bars and filled triangles, respectively.
    We adopt 45\arcdeg as the position angle of the filament for all cores. 
    The magnetic field angles are measured in \citet{Beuther18}.}
    \label{fig:position_angle}
\end{figure}

The molecular outflow axis can be used to infer the rotation axis, and the orientation of outflow axis compared to magnetic field or filament orientation. 
At the core scale, no strong correlation between outflow axis and magnetic field has been reported in both low-mass \citep[e.g.,][]{Hull14,HullZhang19} and high-mass star-forming regions \citep[][]{Zhang14,Baug20}. 
This lack of correlation 
implies that the role of magnetic fields is less important than both gravity and angular momentum from the core to disk scales \citep[e.g.,][]{Sanhueza21}.
A random distribution of outflow-filament orientation has also been found in both low-mass and high-mass star-forming regions \citep[][]{Tatematsu16,Stephens17,Baug20}. 
\citet{Wang11}, \citet{Kong19}, and \citet{Liu20} conducted statistical studies toward the IRDC G28.34+0.06.
They found that outflows are mostly perpendicular to the filament and aligned within 10\arcdeg\ of the core 
-scale ($<$0.05 pc) magnetic field.
\citet{Baug20} found a random orientation of outflows with the filament and the magnetic field in evolved high-mass star-forming region, and argue that its inconsistency with the observation toward IRDC G28.34+0.06 \citep[][]{Wang11,Kong19,Liu20} might come from different evolutionary stages. 
We note that polarization observations toward IRDCs that aim to study magnetic fields are still scarce, with most of the few examples available mostly using single-dish telescopes \citep[][]{Pillai15,Liu18,Soam19}. 

Figure~\ref{fig:position_angle} shows the difference of the projected position angles of outflow ($\theta_\mathrm{outflow}$) with respect to magnetic field orientation ($\theta_\mathrm{B}$) and the filament ($\theta_\mathrm{filament}$), indicating that  outflows are randomly oriented with respect to both the magnetic field and the filament orientation. 
The position angle of the magnetic field was derived from the mode angle in the histogram of polarization orientation angles \citep[Figure~4 in][]{Beuther18} rotated by 90\arcdeg.
We plot the difference between the position angle of the magnetic field and that of the outflow ($|\theta_\mathrm{outflow}-\theta_\mathrm{B}|$) as open squares.
The bar originates from the variation in the histogram of polarization orientation angles.
We adopted 0\arcdeg\ as the magnetic field angle in ALMA4 inferred from visually inspecting Figure~3 in \citet{Beuther18}, though the polarised emission in ALMA4 \citep[mm4 in ][]{Beuther18} is almost unresolved.
As the position angle of the filament, we adopted the position angle ($\theta_\mathrm{filament}\sim$45\arcdeg) of the largest structure identified as `trunk' in the  dendrogram technique. 
The difference ($|\theta_\mathrm{filament}-\theta_\mathrm{outflow}|$) are plotted as filled triangles in Figure~\ref{fig:position_angle}. The angular separations are randomly distributed and no correlation is confirmed.
In our limited sample, we find inconsistent results with respect to what was found in a different IRDC by \citet{Kong19}, which may indicate that the random distribution of the outflows is not due to evolution. 
We should note that this result is affected by the projection effect.
Increasing the number of polarization observations toward IRDCs will certainly help to confirm the recent findings related to the importance of magnetic fields in the early stages of high-mass star formation. 

\subsection{Evolutionary Stages}\label{sec:evolution}

Our ALMA observations unveiled widespread star formation activity in G023.477. 
Detection of CO/SiO outflows and H$_2$CO/CH$_3$OH emission imply the existence of deeply embedded protostars. 
The lines with high upper energy states  ($E_\mathrm{u}>$23\,K) are likely emitted from warm regions which have been heated by embedded protostars. 
Therefore, we use the outflow and high $E_\mathrm{u}$ lines (H$_2$CO $3_{2,1}$--$2_{2,0}$, H$_2$CO $3_{2,2}$--$2_{2,1}$, and CH$_3$OH $4_2$--$3_1$) as star formation signatures. 
Based on these detections, we classified dust cores  into three categories: (i) protostellar cores, (ii) protostellar core candidates, and (iii) prestellar core candidates.
Cores associated with both outflows and 
high excitation lines (ALMA2, ALMA3, ALMA4, and ALMA8) are classified into group (i), 
cores with H$_2$CO or CH$_3$OH emission but no associated to outflows (ALMA1, ALMA5) are categorized as group (ii), and 
cores without H$_2$CO, CH$_3$OH, nor outflows (ALMA6, ALMA7, sub1-3) are classified as group (iii).
Core masses estimated from the 1.3 mm continuum emission (Section~\ref{sec:mass}) range from 1.1 to 19 $M_\odot$ for group (i), 2.3 to 14 $M_\odot$ for group (ii), and 1.4 to 6.4 $M_\odot$ for group (iii). 

As for the group (i) ALMA 2, 3, 4, and 8, the detection of molecular outflows associated to these four cores make them unambiguously protostellar. 
The outflow properties in G023.477 are similar to those in another massive IRDC G28.34+0.06 \citep[][]{Zhang15}.
\citet{Zhang15} also reported that Core 5 in G28.34 is thought to be at a very early phase of evolution and hosts a low-mass protostar by comparing the line spectra with an intermediate-mass protostar in the DR21 filament.
Considering the core mass and the strength of high excitation lines such as H$_2$CO and CH$_3$OH after considering the difference of the beam size, ALMA2 has physically and chemically similar signatures with Core 5. This suggests that ALMA2 has a low-mass protostar at an early phase of evolution, consistent with the short dynamical timescale of $<10^4$\,yr of the outflow.
The relatively higher peak intensity of H$_2$CO and CH$_3$OH, the detection of HC$_3$N ($J$ = 24-23), and the highest rotational temperature suggest that ALMA8 is the most evolved among all cores in G023.477.

ALMA1, identified as mm3 by \citet{Beuther13,Beuther15}, shows a compact structure at 1.3 mm continuum emission in our ALMA data.  
The detection of high excitation lines of H$_2$CO and CH$_3$OH strongly suggest that ALMA1 already hosts an embedded protostar. 
Besides, the rotation temperature estimated from H$_2$CO is the second highest and similar to those measured in the protostellar cores categorized in (i). 
However, ALMA1 is dark even at 100\,$\mu$m \citep[see Figure~2 in][]{Beuther15} and there is no evidence in the current data of an outflow or jet traced by CO or SiO. 
Remarkably, N$_2$D$^+$ has its maximum intensity at the continuum peak of ALMA1. 
The emission of DCO$^+$ is also detected around ALMA1, while that of DCN is relatively weak. 
The C$^{18}$O depletion factor of ALMA1 is higher than those of protostellar cores, group (i), by a factor of $\sim$4. 
These features support that CO sublimation is not yet efficient around ALMA1, implying that the embedded protostellar object has not significantly warmed its surrounding material. 
ALMA1 has a compact continuum emission, the highest density in this region, and no detectable outflows, similar to MM2 in IRDC G11.92-0.61. 
MM2 in G11.92 is a strong dust continuum source without any star formation indicators (no masers, no centimeter continuum, and no (sub)millimeter wavelength line emission including outflow tracers) \citep[][]{Cyganowski14,Cyganowski17}. 
MM2 is a massive ($>30\,M_\odot$) dense
($n_\mathrm{H_2}>10^9$\,cm$^{-3}$ and $N_\mathrm{H_2, peak}>10^{25}$\,cm$^{-2}$) core, and regarded as the best candidate for a bonafide massive prestellar core. 
Comparing ALMA1 with MM2, the detection of some line emission such as H$_2$CO and CH$_3$OH, in addition to strong N$_2$D$^+$ and DCO$^+$, suggests ALMA1 is more chemically evolved.
Thus, ALMA1 seems to be in an extremely early phase of protostellar evolution.

ALMA5 has the lowest rotation temperature and its C$^{18}$O depletion factor is similar to those of protostellar cores. We note, however, that given the position of ALMA5 with respect to the outflows launched from ALMA2 and ALMA3, it is possible that in ALMA5 the detection of H$_2$CO and CH$_3$OH is not internally produced, but externally by the outflow interaction with the core.

ALMA6 and ALMA7 both have no high excitation lines detected (and no outflows), making them  prestellar candidates. ALMA7 has the highest C$^{18}$O depletion factor and the lowest C$^{18}$O column density, suggesting a very cold environment.  

\subsection{Potential for high-mass star formation}\label{sec:highmass}
G023.477 has been regarded as a prestellar, massive clump candidate suitable for the study of the earliest stages of high-mass star formation. 
From previous studies, G023.477 properties are summarized as follows.  The mass and the radius is $\sim$1000 \Msun\ and 0.42 pc, respectively, based on  dust continuum observations \citep{Sridharan05,Yuan17}. The surface and number densities are evaluated as $\sim$0.45 g\,cm$^{-2}$ and $\sim$5.5$\times 10^4$ cm$^{-3}$, respectively \citep[][]{Yuan17}. 
Below, using these global quantities, we discuss whether G023.477 has the potential to form high-mass stars and how high-mass stars can be created in this clump.

The clump surface density is often a good indicator for high-mass star formation. \citet{urquhart14} and \citet{He15} derive an empirical threshold 
for high-mass star formation of 0.05\,gr\,cm$^{-2}$. 
G023.477's surface density significantly exceeds this
threshold. 
Another empirical condition for high-mass star formation is the threshold clump mass derived by \citet{Kauffmann10}. They derive  a mass threshold given as  $M_\mathrm{threshold}=580\,M_\odot (r/\mathrm{pc})^{1.33}$,  by conducting dendrogram analysis of molecular clouds forming low- and high-mass stars. In the case of G023.477, the mass  threshold obtained is $M_\mathrm{threshold}=180\, \,M_\odot$. 
The mass of G023.477 ($\sim$1000\,$M_\odot$)
significantly exceeds this threshold mass. 

Using the observed clump properties, we estimate a possible maximum stellar mass formed in this clump. 
\citet{Larson03} obtain an empirical relation between the total stellar mass of a cluster ($M_\mathrm{cluster}$) and the maximum stellar mass in the cluster ($m^*_\mathrm{max}$) as
\begin{eqnarray}
    m^*_\mathrm{max} &=& 1.2\left( \frac{M_\mathrm{cluster}}{M_\odot}\right)^{0.45}\,M_\odot \\
    &=& 15.6\left( \frac{M_\mathrm{clump}}{10^3\,M_\odot} \frac{\varepsilon_\mathrm{SFE}}{0.3} \right)^{0.45}\,M_\odot,
\end{eqnarray}
where the star formation efficiency, $\epsilon_\mathrm{SFE}$,
is evaluated as $\varepsilon_\mathrm{SFE}=0.1-0.3$
for nearby embedded clusters \citep{Lada03}.
We also assumed the relation of $M_\mathrm{cluster}= \varepsilon_\mathrm{SFE} M_\mathrm{clump}$.
Using the G023.477 clump mass of $10^3\,M_\odot$, the maximum stellar mass derived is 9.5--16\,$M_\odot$.
More recently, using Kroupa's IMF \citep[][]{Kroupa01}, \citet{Sanhueza19} derive another relation for the maximum stellar mass that could be formed in a clump as
\begin{equation}
\label{equ:lada}
    { m^*_\mathrm{max}=\left(\frac{0.3}{\varepsilon_\mathrm{SFE}}\frac{21.0}{M_\mathrm{clump}/M_\odot} + 1.5 \times 10^{-3}\right)^{-0.77}\,M_\odot. }
\end{equation}
From the above equation, the maximum stellar mass is estimated to be 8.3--19\,$M_\odot$.
In summary, the expected maximum 
mass of high-mass stars  formed in G023.477 is estimated to be about 8--19 $M_\odot$ from the empirical relations.

In section~\ref{sec:mass}, we showed that the mass range of the identified cores is from 1.1 to 19 $M_\odot$, which is comparable to the expected maximum stellar mass range.
We should note that there are uncertainties (\ref{sec:mass}) to estimate core masses from dust continuum emission.
We propose two possibilities that may take place in G023.477 to finally form high mass stars from the identified cores: (1) a high star formation efficiency at the core scales and/or (2) additional  accretion onto the cores from the surrounding inter-clump material. We cannot, however, rule out a combination of both possibilities. 

The first one assumes a relatively large star formation efficiency of $\gtrsim$50\%, which, for instance, would enable ALMA2 (19\,$M_\odot$) to form a high-mass star with a mass of $\gtrsim$10\,$M_\odot$ if the core would not fragment into smaller structures.
In this case, no additional mass feeding onto the cores is necessary. This picture is in agreement with the turbulent core accretion scenario \citep{McKeeTan03} and relatively high star formation efficiencies are theoretically possible \citep[e.g.,][]{MatznerMcKee00}.  However, the most massive cores in G023.477 are sub-virialized even after including the magnetic field in the analysis, which is inconsistent with the turbulent core accretion scenario. 

In the second case, the mass feeding would enable the cores to grow and collect the necessary mass to form high-mass stars. Considering the global collapse of the clump suggested by \citet{Beuther15}, the ALMA cores have a large mass reservoir from where to gather additional mass and grow. 
This picture 
is in agreement with competitive accretion scenarios \citep{Bonnell01,Bonnell04}, global hierarchical collapse \citep{vazquez19}, and the inertial flow model \citep{Padoan20, Pelkonen21}.
Recently, \citet{Takemura21} pointed out that the cores need to accumulate gas from their surroundings to reproduce the stellar
IMF from the present core mass function in the Orion Nebula Cluster region.
In a different IRDC of the ASHES survey, \cite{Contreras18} estimate a core infall rate of $2 \times 10^{-3} M_\odot\,\mathrm{yr}^{-1}$. Assuming this infall rate for ALMA1 here in G024.477, in the core free fall time of 7.5 $\times 10^3$\,yr, the core can grow from 15 \Msun\ to a total of 29 \Msun\ and be capable to form a high-mass star.  

Based on our limited case study, we cannot definitely constrain star formation scenarios. However, with a statistical study on the complete ASHES survey and observations of infall tracers, as done in \cite{Contreras18}, we aim to put a firm constraint on theoretical models.

\section{Conclusions} \label{sec:summary}

We have observed IRDC G023.477 at 1.3 mm with ALMA as part of the ASHES survey, obtaining an angular resolution of  $\sim$1$\farcs2$ ($\sim$6000 au in physical scale). G023.477 is a 70 $\mu$m dark IRDC that was previously regarded as a high-mass starless clump with the potential to form high-mass stars. We resolved 11 cores in dust continuum emission and revealed current star formation activity using line emission. The clump can no longer be considered to be prestellar, as it contains cores at very early stages of evolution.  

The 1.3 mm continuum emission unveiled condensed structures embedded in a filament. In addition to the four cores identified in previous works, seven cores are newly detected.
The estimated core masses range from 1.1 $M_\odot$ to 19 $M_\odot$, and the column densities are about 10$^{23}$ cm$^{-2}$. At least four outflows are detected in CO and SiO line emission, indicating star formation has already begun in G023.477 for at least 10$^4$ years. The orientation of outflow axis is randomly oriented compared to the filament and the magnetic field. The PV diagram of the outflows indicates episodic accretion. ALMA3 is the second case of a S-shaped structure in the PV diagram. The detection of high excitation H$_2$CO and CH$_3$OH lines also support active star formation. Based on the detection of outflows and high excitation lines, ALMA1-5 and ALMA8 are  protostellar core candidates. Deuterated molecules trace a slightly different environment, implying ALMA1 is likely to be just after protostellar formation. On the other hand, ALMA8 is the most evolved protostellar core.  
The maximum stellar mass expected in G023.477 is 8--19\,$M_\odot$.
We discuss two possible scenarios in the context of star formation theories under which the IRDC G023.477 would end forming high-mass stars. 

\acknowledgments
K.M is supported by FoPM, WINGS Program, the University of Tokyo. 
P.S. was partially supported by a Grant-in-Aid for Scientific Research (KAKENHI Number 18H01259) of the Japan Society for the Promotion of Science (JSPS). 
S.O. is supported by JSPS KAKENHI Grant Number 20K14533.
H.B. acknowledges support from the European Research Council under the European Community's Horizon 2020 framework program (2014-2020) via the ERC Consolidator Grant `From Cloud to Star Formation (CSF)' (project number 648505). H.B. also acknowledges funding from the Deutsche Forschungsgemeinschaft (DFG) via the Collaborative Research Center (SFB 881) `The Milky Way System' (subproject B1).
Data analysis was in part carried out on the Multi-wavelength Data Analysis System operated by the Astronomy Data Center (ADC), National Astronomical Observatory of Japan. 
This paper uses the following ALMA data: ADS/JAO. ALMA No. 
2018.1.00192.S. ALMA is a partnership of ESO (representing its member states), NSF (USA) and NINS (Japan), together with NRC (Canada), $MOST$ and ASIAA (Taiwan), and KASI (Republic of Korea), in cooperation with the Republic of Chile. The Joint ALMA Observatory is operated by ESO, AUI/NRAO, and NAOJ.
Data analysis was in part carried out on the open use
data analysis computer system at the Astronomy Data
Center (ADC) of the National Astronomical Observatory of Japan.
This research has made use of the NASA/IPAC Infrared Science Archive, which is funded by the National Aeronautics and Space Administration and operated by the California Institute of Technology.
\vspace{5mm}
\facility {ALMA, IRSA, Spitzer, Herschel} 
\software{CASA (v5.4.0,5.6.0; \citealt[][]{McMullin07})}

\bibliography{reference}
\bibliographystyle{aasjournal}

\appendix
\section{Additional figures}
\begin{figure}
    \centering
    \epsscale{0.8}
    \plotone{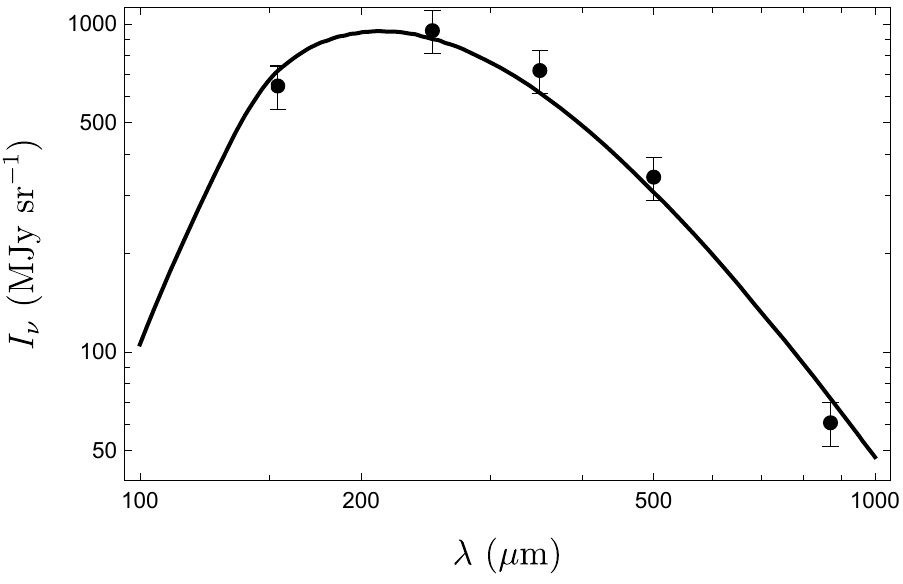}
    \caption{SED fittings performed over the intensities measured at the peak of the 870 $\mu$m image.  The estimated dust temperature is $13.8 \pm 0.8$ K.}
    \label{fig:SED}
\end{figure}

\begin{figure}
    \centering
    \plotone{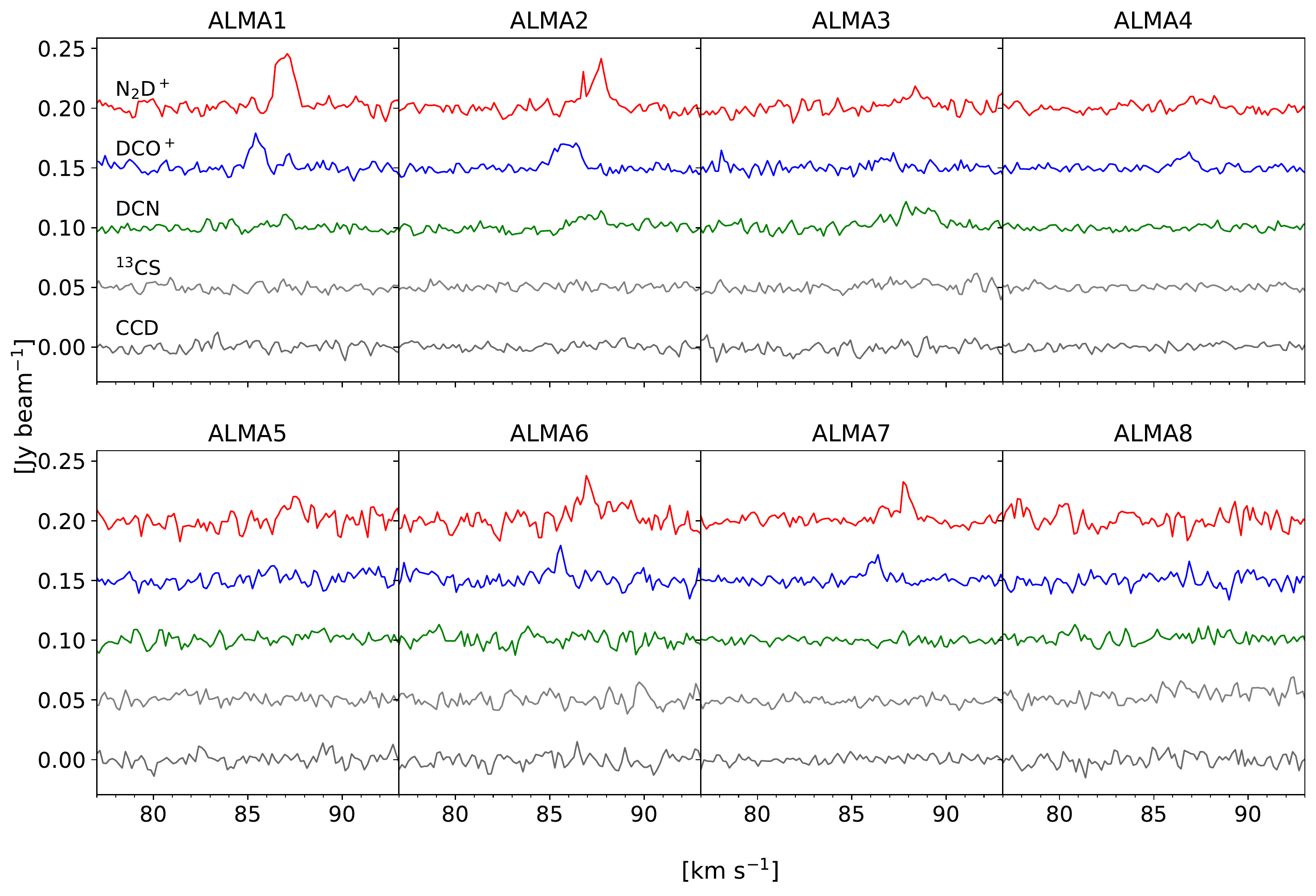}
    \caption{Core-averaged spectra of N$_2$D$^+$ ($J$=3–-2) (red), DCO$+$ ($J$=3-–2) (blue), DCN ($J$=3-–2) (green), $^{13}$CS ($J$=5--4) (gray), and CCD ($N$=3--2) (dark gray) of ALMA1--ALMA8. These spectra are averaged within core areas identified by the dendrogram algorithm.}
    \label{fig:ave_profile}
\end{figure}

\begin{figure}
    \centering
    \plotone{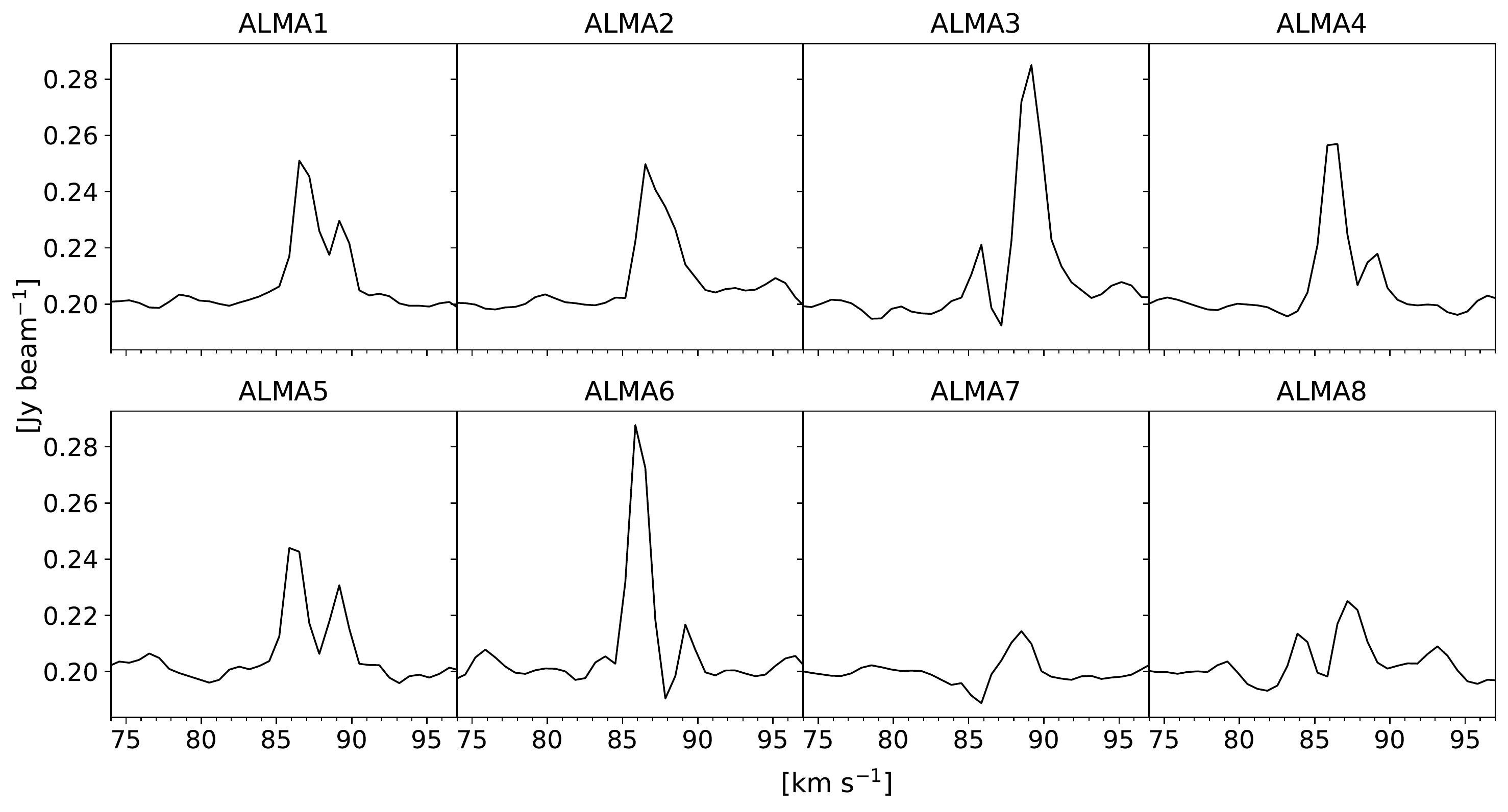}
    \caption{Core-averaged spectra of C$^{18}$O ($J$=2--1) of ALMA1--ALMA8. These spectra are averaged within core areas identified by the dendrogram algorithm.}
    \label{fig:C18O_profile}
\end{figure}

\begin{figure}
    \centering
    \plotone{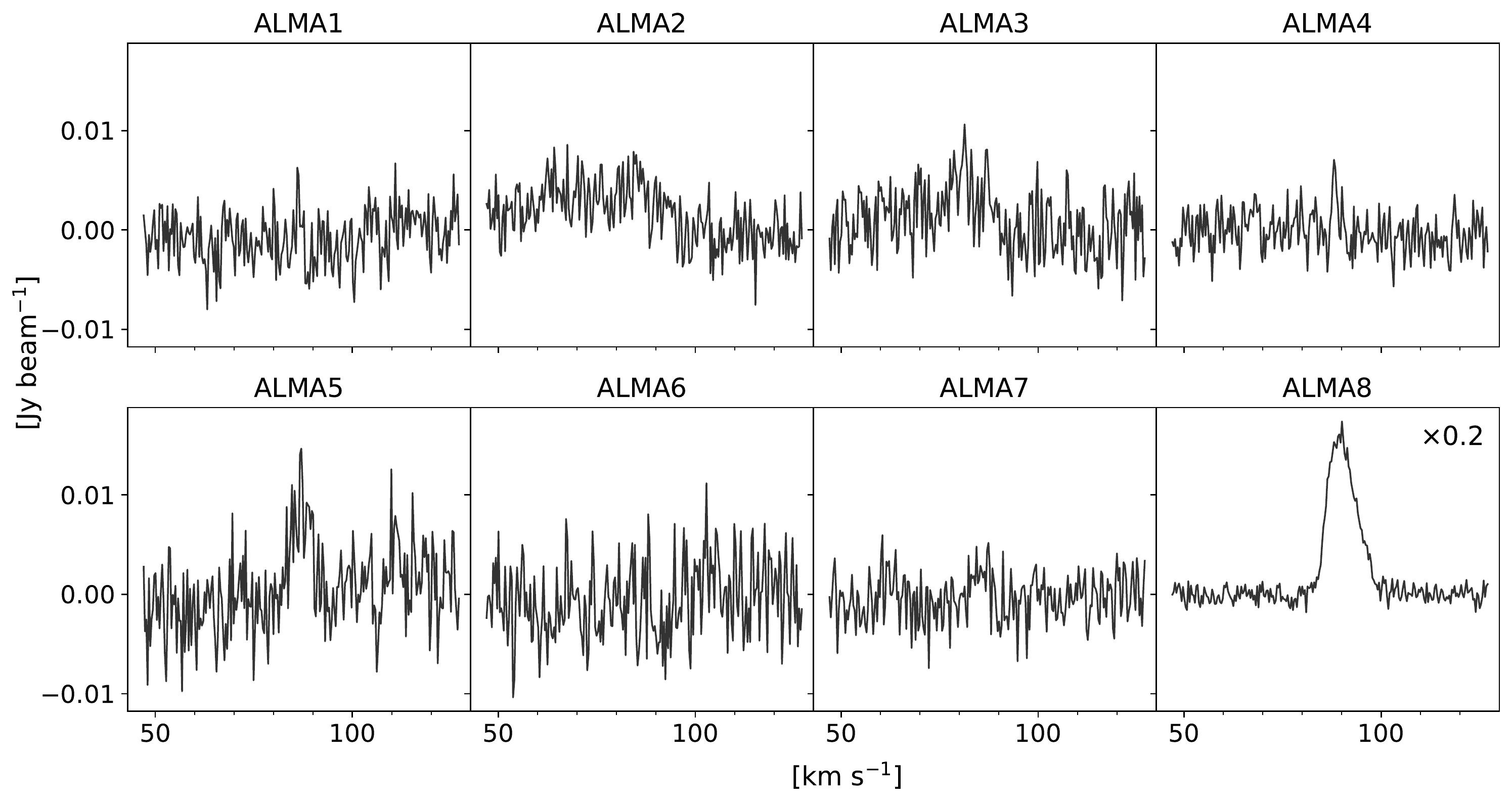}
    \caption{Core-averaged spectra of SiO ($J$=5--4) of ALMA1--ALMA8. The intensity of ALMA8 is plotted multiplied by 0.2.}
    \label{fig:SiO_profile}
\end{figure}

\begin{figure}
    \centering
    \epsscale{1.05}
    \plotone{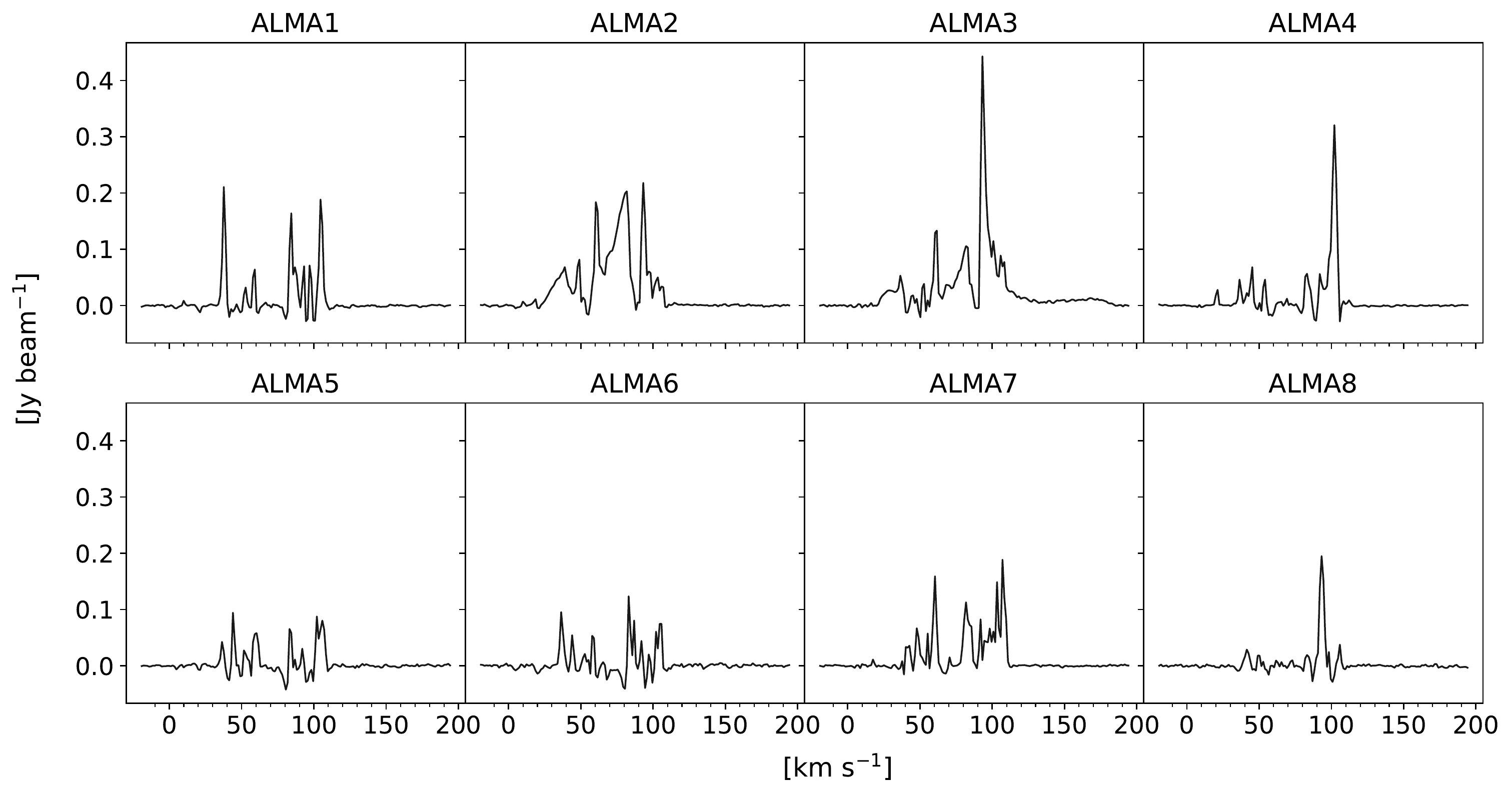}
    \caption{Core-averaged spectra of CO ($J$=2--1) of ALMA1--ALMA8.}
    \label{fig:CO_profile}
\end{figure}

\begin{figure*}
    \plotone{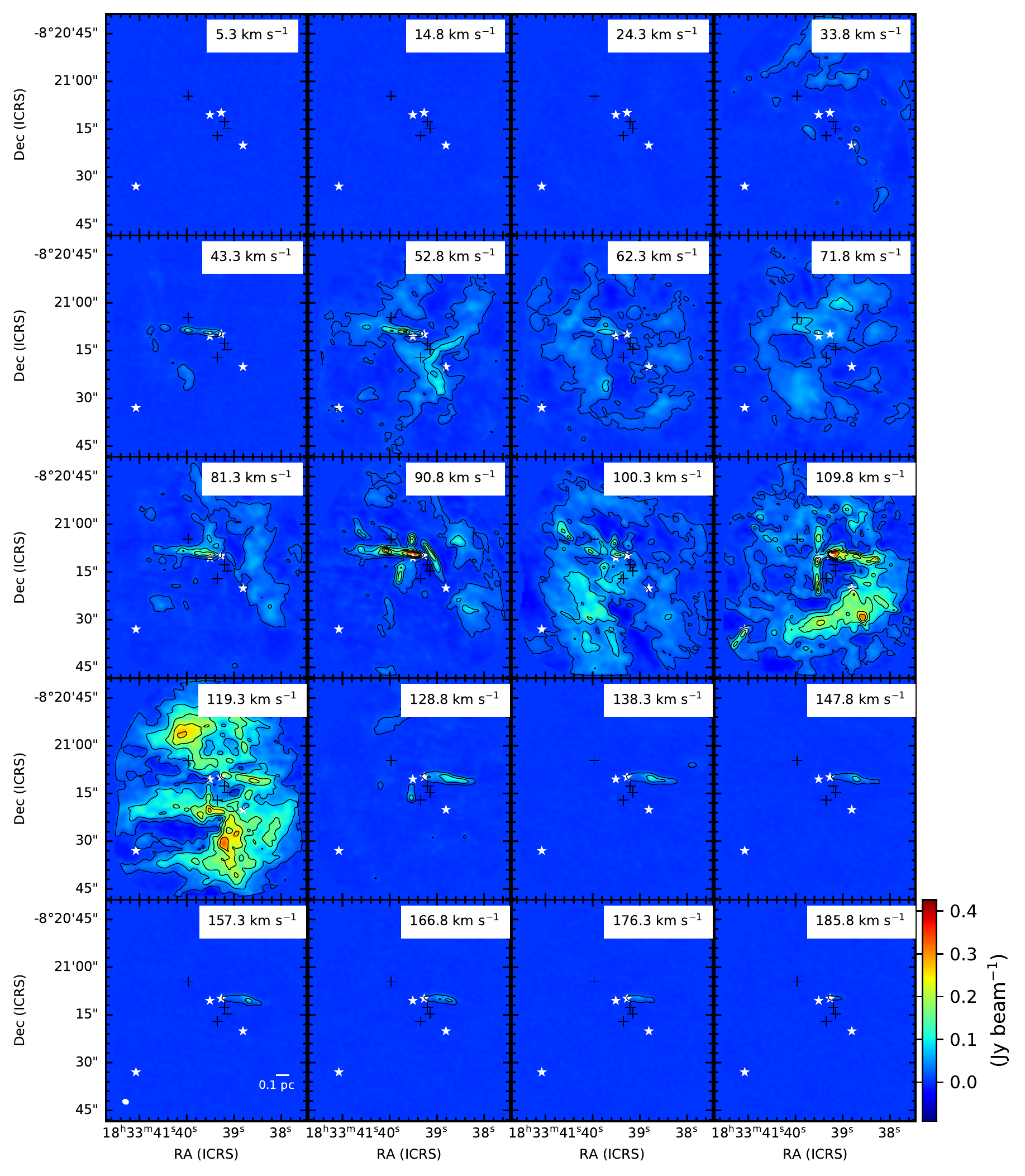}
    \caption{Channel maps of CO ($J$=2--1). 
    The contour levels are 10, 50, 100, 150, and 200$\sigma$ (1$\sigma$=\,1.35\,mJy\,beam$^{-1}$).  The white star symbols represent the continuum peak positions of cores (ALMA2, 3, 4, and 8) associated with outflows. The plus symbols represent the continuum peak position of ALMA1, 5, 6, and 7 (no outflow).
    The spatial scale and the beam size are shown at the bottom.} 
    \label{fig:channelmapCO}
\end{figure*}

\begin{figure*}
    \plotone{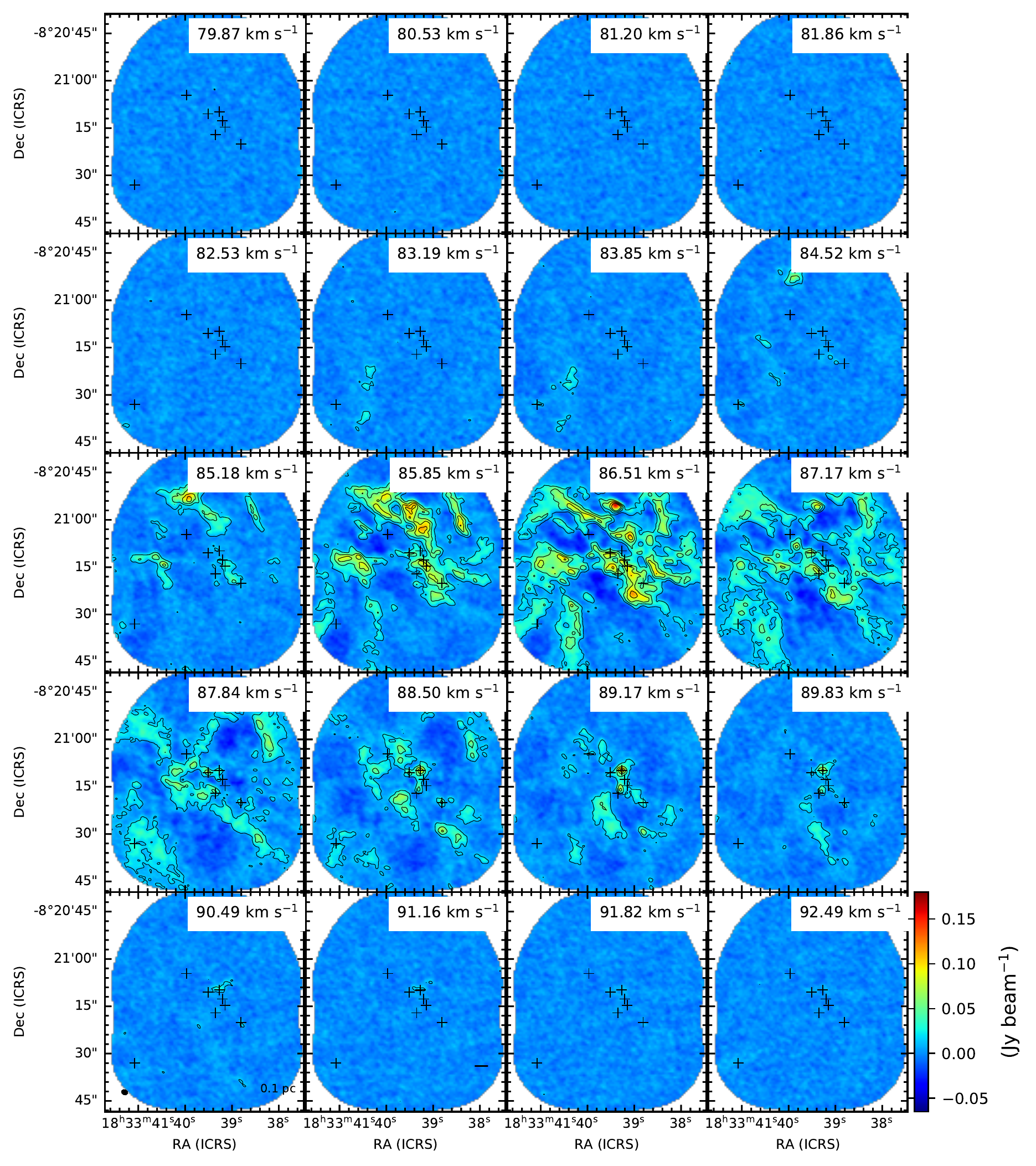}
    \caption{Channel maps of C$^{18}$O ($J$=2--1).
    The contour levels are 4, 10, 20, and 30$\sigma$ (1$\sigma$=\,3.73\,mJy\,beam$^{-1}$). The plus symbols represent the continuum peak position of ALMA1-ALMA8.
    The spatial scale and the beam size are shown at the bottom.}
    \label{fig:channelC18O}
\end{figure*}

\begin{figure*}
    \plotone{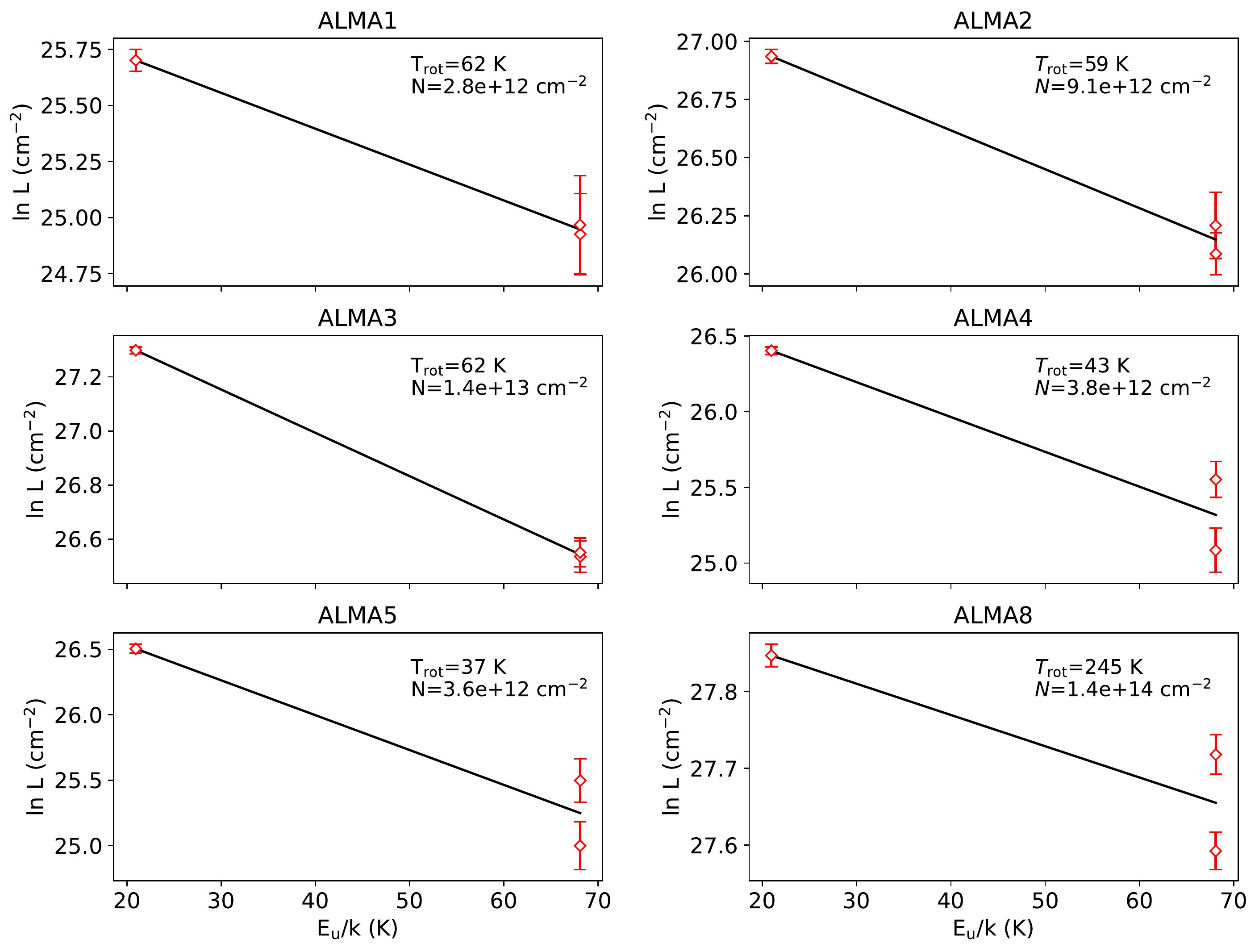}
    \caption{H$_2$CO rotational diagrams.
    Red points represent observational measurements and the black lines are the fitting results. The derived parameters rotational temperature and column density of H$_2$CO are shown on the right top on each panel and summarized in Table~\ref{tab:phypara}. The error bars correspond to 1$\sigma$ uncertainties.}
    \label{fig:rot_diagram}
\end{figure*}

Figure~\ref{fig:SED} shows the result of SED fitting. The derived temperature was used in estimating core mass and the C$^{18}$O depletion factors. The exact measured fluxes are described in Section~\ref{sec:mass}.
Spectra averaged core areas (ALMA1--ALMA8) of dense gas tracers, C$^{18}$O, SiO, and CO are summarized in Figure~\ref{fig:ave_profile}, \ref{fig:C18O_profile}, \ref{fig:SiO_profile}, and \ref{fig:CO_profile}, respectively. They are averaged within core areas which dendrogram identified.
Figure~\ref{fig:channelmapCO} and \ref{fig:channelC18O} are the channel map of CO and C$^{18}$O emission. Figure~\ref{fig:rot_diagram} shows the H$_2$CO rotation diagram, which is used to estimate the rotational temperature and the column density of H$_2$CO in Section~\ref{sec:high-ex}.




\end{document}